# Incorporation of Strong Motion Duration in Incremental-based Seismic Assessments

Mohammadreza Mashayekhi[1]*, Mojtaba Harati[2], Atefe Darzi[3], and Homayoon E. Estekanchi[1]

[1]Department of Civil Engineering, Sharif University of Technology, Tehran, Iran
[2]Department of Civil Engineering, University of Science and Culture, Rasht, Iran
[3]Faculty of Civil and Environmental Engineering, University of Iceland, Reykjavik, Iceland

**Abstract**

This study proposes a new approach to incorporate motion duration in incremental dynamic assessments. In the proposed methodology, at each intensity level, a simulation-based approach—which is verified with actual data—is employed to determine the median duration and the median acceleration spectra of ground motions expected to occur at the site. Afterward, at each intensity level, artificial or spectrally matched motions are produced based on the median acceleration spectra and the median duration, indicating that different intensity levels are directly covered by the generated artificial or adjusted motions rather than just scaling up and down a set of recorded ground motions. In the proposed methodology, duration and acceleration spectral shape changes against intensity level while they remain the same for different intensity levels in approach where responses are derived by scaling up and down of a set of ground motions. The functional relationship between duration and seismic intensity level, which is vital for the estimation of median duration at each intensity level, is firstly investigated for the sites with different soil conditions and rupture distances. Not only is it demonstrated that the data can fit into exponential functions, but the sensitivity of the functions against different parameters is also explored as well. The proposed duration-consistent incremental seismic assessment is used in nonlinear seismic assessment of two single degree of freedom structures—with and without a degrading behavior capability. It is revealed that when changes in duration and spectral shape of the motions at different intensity levels are considered in the nonlinear dynamic analysis, an impactful influence that cannot be easily ignored is witnessed in the structural responses of incremental analyses.

## 1. Introduction

Three features characterize the destructiveness power of strong ground motions—intensity, frequency contents and the duration. All these three features must be incorporated in dynamic analysis of structures. Amplitude-based intensity measures include Peak Ground Acceleration (PGA), Peak Ground Velocity (PGV) and Spectral Acceleration (SA) of the recorded ground motions. Frequency content of earthquake motions are extracted by Fourier spectrum or wavelet analysis but it is generally accepted that frequency content is reflected in spectral acceleration. In fact, considering spectral acceleration implies that both intensity and frequency content are considered in analysis.

Amplitude-based properties are the most important and primary ground motion parameters of the earthquakes, which are found to have a good correlation with structural damages imposed on the built infrastructures after severe earthquakes. For this reason, the PGA and SA at a specific period of vibration are usually taken as the main amplitude-based IM parameter for the record selection procedures of the dynamic time-history analyses (Iervolino and Cornell 2005, Araújo et al. 2016).

---

* **Corresponding author:** Mohammadreza Mashayekhi, Department of Civil Engineering, Sharif University of Technology, Tehran, Iran. e-mail: mmashayekhi67@gmail.com

Even though it was shown that duration can play an important role in the response of structures under earthquake motions, this parameter has not been directly included in the structural analysis (Bommer et al. 2004). It was reported that seismic responses of the structures under earthquake loadings with deteriorative behaviors, including RC frames (Belejo et al. 2017, Chandramohan et al. 2016, Han et al. 2017, Hancock and Bommer 2007, Raghunandan and Liel 2013), concrete dams (Sherong et al. 2013, Wang et al. 2015, Xu et al. 2018, Wang et al. 2018), SDOF systems with pinching-degrading behavior (Molazadeh and Saffari 2018) and masonry buildings (Bommer et al. 2004), are much more sensitive to the duration of the ground motions. It means that more damages would be expected to occur in the structures that are subjected to long-duration ground shakings (Harati et al. 2019, Mashayekhi et al. 2019b). In this regard, it has been found that motion duration metrics of the earthquake records, which are not being included in the traditional dynamic analysis procedure now, may have more than 80% correlation with the cumulative damage indices (Guo et al. 2018, Mashayekhi et al. 2019). It is worth mentioning that in these investigations, using a spectral matching procedure, the motion duration of the selected GMs was isolated from the other characteristics of the employed earthquake records. In a spectral matching procedure like the one proposed by Hong and Huang (2020), GM records are adjusted in a way that their response spectra get matched to a target acceleration spectrum. In this way, the amplitude-based intensity IMs of the records, including the SA, spectral velocity (SV) and spectral displacement (SD), would get unified and become quite close to the spectrum-based IMs of the target response spectra. Consequently, the final modified GMs would only differ in terms of motion duration and the non-stationary characteristics properties they inherit from the original or unmatched earthquake records. Finally, these modified ground motions are utilized for the nonlinear dynamic analyses of the considered structures in order to find a correlation between motion duration and a damage metric, e.g. a DM such as the Pak-Ang index. In this case, it is reported that accumulated damage indices, which are based on hysteretic cyclic energy of the earthquakes such as the Pak-Ang damage index, have been found to have higher correlations with motion duration. Although it was demonstrated that the extreme damage indices such as peak floor drifts or peak plastic rotations of the elements are not highly dependent to motion duration, it has been found that they are also susceptible to motion duration of GMs (Hancock and Bommer 2007, Sarieddine and Lin 2013, Mashayekhi et al. 2019c). It should be noted that the same results as mentioned for deteriorative structures apply for the steel (Bravo-Haro and Elghazouli 2018, Hammad and Moustafa (2019, 2020)) and wood frame (Pan et al. 2018) structures.

In order to include the duration in structural analyses, it is required to quantify the motion duration of the earthquakes. Several duration definitions are available in the literature, but none of them has been unanimously accepted as a reliable duration indicator by earthquake engineers (Bommer et al. 2009). All these duration definitions are categorized into two different groups, the direct and indirect definitions. Direct detentions measure the duration of earthquake motions based on the strong ground motion, while indirect definitions calculate the duration based on the response of structures subjected to the ground motion. Three types of direct definitions are available—bracketed, uniform and significant duration. The bracketed duration of motion specifies the total time left between the first and last acceleration excursions which are greater than a specific predefined threshold (0.05g for instance). Uniform duration measures the total time of the earthquake motion in which the amplitudes exceed a specified threshold. (Bommer and Marytínezpereira 1999). Significant duration is denoted by Dx-y herein, which is defined as the time interval during which the normalized Arias Intensity (AI) goes up from a minimum (x%) to a maximum (y%) threshold. The formula of calculating Arias Intensity is given in Equation (1).

$$AI = \frac{\pi}{2g} \int_{0}^{t_{max}} \left[ a(t) \right]^2 dt \qquad (1)$$

where the $|a(t)|$ is the absolute value of the acceleration function of the ground motion at time $t$, a(t). Also $t_{max}$ and AI are the total duration of ground motion and the total AI calculated for the entire duration of the ground shakings. It can be readily understood from the given form of the above-mentioned equation that this duration-related intensity measure is in continuous form and increases with time and has the capacity to capture the accumulative characteristics of the earthquakes (Mashayekhi et al. 2018). As an example in this case, the procedure pertinent to the calculation of the D5-75 parameter for the Loma-Prieta earthquake of 1989 is depicted in Figure 1. According to this figure, significant duration or D5-75 is the time interval during which the buildup energy of the AI jumps from a minimum (5%) to a maximum (75%) threshold level. The times associated with the mentioned minimum and maximum thresholds in the time history profile of this GM are defined by $t_x$ (here 8sec) and $t_y$ (here 12.8sec), respectively.

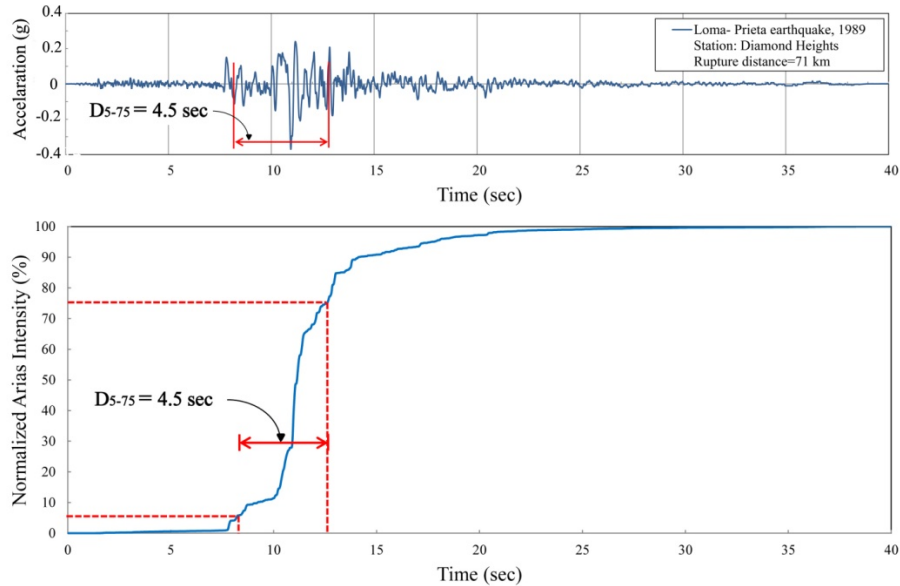

Figure 1. The procedure required to compute the D5-75 parameter of a recorded ground motion

In order to include the duration in time history analysis, the relationship between duration and amplitude-based intensity measures must be determined. Different characteristics of the earthquakes—such as duration and amplitude-based IMs of the GMs—are simultaneously shaped and originated from the same source, site and path conditions events move from a source to a specific site. Hence, the role of a mechanism for the motion duration updating at different seismic intensity levels should get much consideration for study as a potential influencing factor for the record selection procedures of dynamic frameworks (Shoji et al. 2005).

A number of researches have been conducted to explore the possible relationship or correlation of duration-related intensity measures with SA, PGA or PGV for example as the amplitude-based intensity measures that are practically used as the main record selection criteria. Bradley (2015) used a combination of ground motion prediction equations (GMPEs) and the bootstrap sampling method to obtain linear correlation coefficients for different intensity measures. He points out that a good positive correlation exists between SA and AI—not a significant duration—over a range of short structural periods while these two considered parameters are not well correlated in periods elsewhere. Nearly the same results for the correlation of SA and Cumulative Absolute Velocity (CAV)—a duration-related velocity like AI—have been reported by Wang and Du (2012) and Bradley (2011), which were obtained using the NGA ground motion database and some adopted GMPEs, respectively. Bradley (2012) and Baker and Bradley (2017) computed linear correlation coefficients for finding such correlations from observed data of the NGA projects. They found that significant duration is negatively correlated with SA in points located in the range of short to medium structural periods although a positive correlation can be witnessed in long periods of vibration. Shoji et al. (2005) also examined the relationship of significant duration and amplitude-based IMs of earthquake motions. They used the hypocentral distance of the events as an intermediary variable to probe the potential relationship of PGA and motion duration of the considered GMs. It is reported that motion duration and PGA have a reciprocal relationship against the hypocentral distance of the events. This means that while the PGA tends to decay at farther hypocentral distances, the motion duration of the events has a tendency to be enlarged at the same time. Similarly, Wang et al. (2002) did the same procedure for finding a possible relationship between bracketed duration and amplitude-based characteristics of the near-fault earthquakes. Using closest distance to fault as an intermediary variable, they saw a direct descending pattern between duration and PGA for the near-fault events recorded in 1999 Chi-Chi, Taiwan Earthquake. The contradictory result of this study—when compared to the findings reported by Shoji et al. (2005)—seems to be pertinent to the fact that the type of considered earthquakes and definition of motion duration is different from the ones chosen by Shoji et al. (2005). The correlation of SA and the effective number of nonlinear cycles as a duration-related intensity measure has been also investigated by yet another research (Du and Wang 2017). It was reported that these two parameters have a relatively good correlation in short structural periods while they experience a descending trend for the correlation coefficient values calculated for the medium to long periods of vibration. It should be mentioned that the data used by the current aforementioned reviewed studies on the relationship of duration and the amplitude-based measures are from dissimilar sets of data, meaning that they are not from the same influencing seismic factors, i.e. from the same source, site and path conditions for both duration and amplitude-based IMs. Therefore, to sync duration to the amplitude properties of GMs in incremental seismic assessments, we need to seek such relationships that are based on the data coming from the same influencing and controlling seismic factors.

After the development of the relationship between amplitude-based intensity measures and duration, this relationship function must be included in the procedure of the dynamic analysis. One of the most well-known dynamic analysis is Incremental Dynamic Analysis (IDA) (Vamvatsikos and Cornell 2002). The IDA is a parametric analysis method for the estimation of seismic demand and capacity for different levels of seismic intensity measures (IMs). In a single-record IDA, an unscaled earthquake record is first selected from an authorized GM database. To track structural responses at multiple levels of excitation, the PGA or the SA of the selected ground motion (GM) is then scaled up and down to cover all desired levels of structural behavior from an elastic status to more plastic ranges, leading eventually to a point recognized as a global collapse capacity of the system (Vamvatsikos and Cornell 2002). Thus, multiple runs of Nonlinear Dynamic Procedure have to be executed in this case to cover all the IMs required to see a continuous status of the structural responses under scaled GMs. In this way, the intensity level of the selected record is changed by a simple transformation which can increase or decrease amplitudes of the motion selected for the single-record IDA study. In other words, a number of scaled duplicates of an accelerogram with different intensity levels are produced from the original (as-recorded or unscaled) accelerogram (Vamvatsikos and Cornell 2002; Mackie and Stojadinovic 2005). Any desirable damage measure (DM) can be hired to record the structural responses through multiple levels of IM. The DMs computed at different levels of IM are normally plotted against the respective level of IMs employed. And, therefore, one continuous response curve—known as the single IDA curve—is obtained in this way. For performing a multi-record IDA study, a set of GMs should be selected in advance for the entire IDA procedure. In this way, the dispersion of the responses at a given level of IM is no longer of deterministic type and can be represented using conditional probability function, $DM = f(IM)$ (Vamvatsikos and Cornell 2002). To date, a multi-record IDA study can be considered as the most reliable and comprehensive dynamic analytical procedure, which is capable of computing structural responses of the built infrastructures at multiple seismic levels up to where the collapsing mechanism of the considered structural systems would be appropriately examined under several candidate GM records.

While the multi-record IDA framework is now being considered as the most reliable computational tool for estimation of seismic demands of structures, researchers are trying to offer some ways to make this framework become less computationally demanding or sound more realistic. For this purpose, few modifications to the procedure of an IDA framework are suggested in such studies. Since the fact that an IDA study needs numerous NDP analyses at multiple seismic levels from various selected GMs, it can be understood that it demands massive computational efforts. In this case, to reduce or obviate the computational demands of an IDA study, one may decide to utilize the recommended equivalent SDOF systems (Vamvatsikos and Cornell 2005), or use a type of pushover analyses on the detailed 3-D MDOF models to predict the IDA responses (Han and Chopra 2006; Soleimani et al. 2018). Moreover, record selection procedures by which a reduced number of selected GMs would be required for a reliable estimation of median IDA response have been also proposed by several researchers (Azarbakht and Dolˇsek 2007; Mashayekhi et al. 2019a). Instead of using a record set for the entire analysis in the IDA, Lin and Baker (2013) introduced a new IDA procedure—named Adaptive Incremental Dynamic Analysis (AIDA)—in which a new record selection is conducted at each level of IM. In an effort to make the IDA become more realistic, the framework of AIDA becomes in fact more consistent with the concept of GM deaggregation in the Probabilistic Seismic Hazard Analysis (PSHA). In this way, target distributions of the significant characteristics of the employed GMs at each level of IM change once the level of IM in the AIDA gets altered. In terms of suggested record selection procedures, the focus of investigations was put on improving or optimizing the record selection methods that are based on the amplitude-based parameters of earthquake records. But the role of motion duration as a potentially important factor of GMs is not yet being considered or has studied enough for the record selection procedures of current dynamic frameworks that include standard and traditional IDA practices as well.

A simulation-based methodology is proposed in this paper to incorporate the influence of motion duration on the structural earthquake demands, thereby trying to offer a more accurate and duration-dependent analytical framework for incremental-based dynamic analyses. In this framework, first a target acceleration spectrum and a target motion duration are generated and simulated at each level of IM that our proposed dynamic analysis is based on. This procedure should be separately conducted for producing a pair of targets, on acceleration spectrum and motion duration, at all IM levels. It is needless to say that for performing local site investigations (Mase et al. 2018, Mase et al. 2018b) or examining the seismic risk of the built infrastructures (Baker et al. 2006), one may need to conduct an assessment on the rate of occurrence of the potential hazard of the earthquake ground motions and tracking the possible effects of these motions on the dynamic responses of the considered components. These two quantities can be typically linked to each other using an IM metric such as spectral acceleration which is normally reflected in the target acceleration spectra of a particular region. So as a next step in this proposed dynamic framework, duration-consistent ground motions are generated by modifying the selected initial motions in such a way that they get fitted to the target acceleration spectra and target durations both of which are derived by a simulation method at different seismic intensity levels. These duration-dependent motions that are generated at all IM levels—from the proposed simulation procedure—are then used to compute the structural dynamic responses for the entire analysis. Since ground motion parameters such as spectral acceleration and motion duration expected at a specific site can be well estimated through a seismic hazard analysis that works based on the developed attenuation models (Anbazhagan et al. 2009, Mase et al. 2018a), the ground-motion models—which are themselves according to the real ground motion data—have been similarly decided to be utilized in this study for the proposed simulation process. To show the efficiency of the proposed method, numerical examples associated with the validation of the simulation method and application of the proposed methodology in the nonlinear dynamic assessment are thoroughly presented and discussed. To finish, a discussion along with a conclusion section is provided to deliver the main findings of this investigation.

## 2. Methodology

In the conventional incremental dynamic analysis, various levels of IM are produced by scaling up and down of a set of GM suite. In the IDA, as mentioned before, this scaling procedure is utilized as a mechanism to change the level of IM at which dynamic time-history analysis should be performed for a set of GMs. However, this scaling process leads to the same spectral shape and duration for each individual earthquake record at different seismic levels considered. On the contrary, the proposed method is developed in a way that it incorporates the variation of spectral shape and motion duration with respect to the chosen level of IM on which dynamic time-history analysis procedures should be performed.

In contrast to IDA framework, levels of IM are not considered to be increased with the GM record scaling in the proposed framework; the intensity level would be intensified with a devised simulation-based mechanism in this recommended dynamic procedure while both the spectral shape and motion duration of the employed earthquake records get changed for all selected or required levels of IM. At each of the levels considered for the IM metric in the proposed method, a unique acceleration spectrum and a value for the median motion duration of the potential GMs would be estimated through a simulation procedure. In fact, target duration and target acceleration spectrum at each level of IM is determined by a simulation approach that is verified by actual data and works based on ground-motion models (e.g. Darzi et al. 2019, Boore et al. 2014). It is worth adding that the lack of enough recorded data for a given site justifies the use of the simulation approach. With these two targets in hand for each level of IM, duration consistent artificial ground motions are generated by adjusting random initial motions so that they get matched to a unique target acceleration spectrum and a specific target duration. However, modified or spectrally matched GMs can be also employed at each level of IM if the initial (or unmatched) motions are selected based on the predicted unique acceleration spectrum and the median value of motion duration estimated at the considered IM level. Since the motion duration consistency can be considered in generating artificial or matched motions, the earthquake duration of the employed GMs would be directly incorporated in the structural analysis. The methodology is presented here in a stepwise manner as follow:

- Select an appropriate IM, e.g. peak ground acceleration, peak ground velocity or acceleration spectra at first mode period. Afterward, specify a range for the IM required for an IDA procedure and then discretize the considered IM over this range at $N_{IM}$ points. The arithmetic discretization of IM levels is shown through Equation (2)

$$IM_i = IM_{\min} + i \times \frac{(IM_{\max} - IM_{\min})}{N_{IM}} \tag{2}$$

where *IM<sub>min</sub>* and *IM<sub>max</sub>* are, respectively, the minimum and maximum considered *IM* values; $N_{IM}$ is the number of discretization points, and *IM<sub>i</sub>* is the *i*-th level of IM in the aforementioned range of IMs that one considers using for the proposed incremental-based analytical method. So, after the completion of this step, we can readily find a candidate value of increment for the *IM*, e.g. 0.05g in a SA scale, by which the proposed method can be executed up to the selected *IM<sub>max</sub>*.

- Choose a duration definition for the proposed dynamic analysis. To select a suitable metric to quantify the motion duration for the proposed dynamic analysis, one may think that there are several definitions to pick for motion duration. But the duration metric selected in this step should be consistent and compatible with the simulation procedure employed for the data sampling. In this regard, we have selected the D5-75 definition of the significant duration that is compatible with the simulation procedure we described in the next section.

- Develop and create relationship functions for spectral acceleration (or target acceleration spectrum) and motion duration as a function of IM level. As mentioned earlier, these relationship functions are obtained using the events' data provided through a simulation process. Please refer to Section 3 for the comprehensive explanations regarding the derivation of these relationship functions. General forms of these functions can be written as:

$$dur = f_{Dur}(IM) \qquad (3)$$
$$S_a = g_{Spec}(T, IM) \qquad (4)$$

where the functions *f (IM)* and *g (T, IM)* are random lines (or random functions), which are produced by the simulation procedure to display the variability of motion duration and spectral shape of a set of hypothesized GMs that would be considered for a given level of IM in the proposed dynamic procedure. As a result, a unique target acceleration spectrum and a distinct value for the median motion duration of the probable GMs can be eventually predicated for each level of IMs that have been selected for this proposed method in the first step.

- Generate $N_{GM}$ artificial or modified (spectrally matched) motions at each IM level based on the information derived in the previous step. In fact, the GMs at each level of IM can be simulated or modified using the target quantities—the target acceleration spectrum and motion duration estimated in the previous step. Accordingly, $N_{GM} \times N_{IM}$ artificial or adjusted ground motions would be produced at this stage as it is indicated in Equation (5):

$$GM_{i,j} = GM\left(IM_i, S_a = g_{spec}(T, IM_i), dur = f_{Dur}(IM_i), X_j\right) \qquad (5)$$

where *j* is an index representing the required number of GMs that should be created at each level of IM, the *IM<sub>i</sub>* in Equation (2). And *X<sub>j</sub>* is the initial motion used to simulate j-th ground motions, where this initial motion is randomly simulated if artificial GMs are intended to represent the earthquake inputs at each level of IM. It is essential to declare that the initial motion (*X<sub>j</sub>*) can be also pick out from existing recorded earthquake motions provided that they are selected based on the targets, the generated acceleration spectrum and motion duration, at each level of IM.

In order to quantify the expected misfit of the adjusted or artificially generated signals, Equation (6) that is so similar to its discrete form—as recommended by Hancock et al. (2006)—is presented here. In this equation, the average misfit of the signal with a target acceleration spectrum at a desired damping ratio is expressed in percent.

$$M(\xi) = \frac{\int_{T_{min}}^{T_{max}} (S_a(T,\xi) - S_{aT}(T,\xi)) \, dT}{T_{max} - T_{min}} \tag{6}$$

where $S_a(T,\xi)$ and $S_{aT}(T,\xi)$ are, respectively, acceleration spectrum of a generated signal and the target spectrum at period $T$ with a damping ratio of $\xi$. The $T_{min}$ and $T_{max}$ are the minimum and the maximum periods between which the average misfit of the generated signal is intended to be computed.

- Perform nonlinear time-history dynamic analysis of the structure under all simulated or modified ground motions. In fact, since $N_{GM}$ number of GMs are generated at each IM level in the previous step, $N_{GM}$ number of nonlinear time-history analyses should be subsequently conducted at each level IM. And this procedure should be repeated for all IM levels. In this step, all the critical response parameters such as inter-story drift ratio or energy-based damage measures (DMs) should be also recorded during the time history analyses.

- Evaluating and processing the engineering demand parameters (EDPs) of the structure for the dynamic analyses performed in all IM levels. One possible engineering demand parameter is the maximum inter-story drift ratio—as mentioned before—though one can also choose DMs such as the Park-Ang metric (Park et al. 1987) as a candidate choice for the nominated EDPs. The matrix of EDP, as presented in Equation (7), would be the final output of this step.

$$EDP_{ij} = \left[ EDP(GM_{ij}) \right] \tag{7}$$

- Plot each individual pairs of IM versus EDP values, i.e. ($IM_i$, $EDP_{ij}$), to show the scattering trend of the results. Indeed, the IM-versus-EDP pairs should be formed at each level of IM. When these outputs are plotted for all levels of IM, the scattering data of the structural responses can be then produced for all the IMs considered.

- Calculate the median and the standard deviation of EDP values at each level of IM. These central values of EDP—for example, the medians or the mean values as brought in Equation (8)—should be determined for all considered levels of IM. Plot pairs of central EDP values against the discretized IMs, i.e. ($IM_i$, $EDP_i$), to obtain a duration-dependent response curve of a structure analyzed by the proposed method.

$$EDP_i = \frac{\sum_{j=1}^{N_{GM}} EDP_{ij}}{N_{GM}} \tag{8}$$

The logic of the proposed methodology is illustrated in Figure 2.

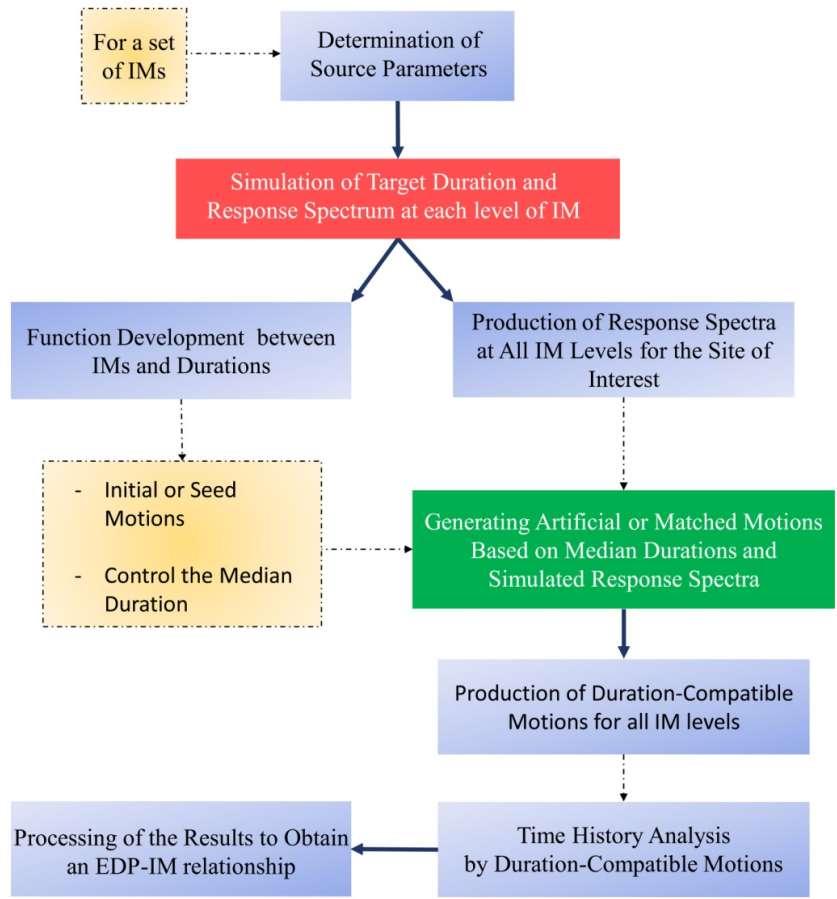

Figure 2. a flowchart representing the methodology of the proposed duration-dependent dynamic analysis

## 3. Relationship between motion duration and intensity measures

### 3.1. Proposed simulation procedure

In this section, a procedure for determining the relationship function between a candidate IM and motion duration is explained for a specific site. The resulting relationship functions as mentioned in Section 2, are used in the method developed in this study—the duration-consistent incremental dynamic analysis. For establishing such a relationship function, Monte Carlo Simulation (MCS) is employed to be at work for the data sampling that is explained below. MCS is a powerful statistical analysis method that is widely applied to many problems existing in different fields to perform numerous experiments on the computer; it can be used in complex and highly nonlinear engineering models because a lot of random variables with different distribution types can be handled without difficulty in this simulation framework. It is worth mentioning that the MCS is being employed frequently in the seismic risk assessment as well (e.g. Bourne et al. 2015, Bommer et al. 2017). For each experiment in the MCS, a set of input random variables $X = (X_1, X_2, ..., X_n)$ is sampled or generated. Then the output variable or the performance function $Y = g(X)$ is computed using the input data of all rounds of the experiment. The following four steps should be performed for reaching out a relationship function between motion duration and the IM selected to be employed for proposed duration-sensitive dynamic analysis:

   i.  First, we should find characteristics of the earthquakes that can all produce a specific level of IM. For this purpose, we have to use ground motion prediction equations (GMPEs). These relationships are typically dependent on a number of contributing elements which include the parameters pertinent to the site, source, and variables standing for the distance-related regressors. In this regard, an appropriate GMPE giving predictions for the amplitude-based characteristics of the earthquake events should be selected for converting the chosen IMs such as PGA and SA—as it is discretized in Section 2—to the characteristics of the expected earthquake records. It is worthwhile to add that the usage of a GMPE in such inverse problem-solving cases is in contrast to the conventional application of such ground-motion models. In this study, the GMPE of Boore et al. (2014) has been selected for the above-mentioned inverse problem-solving process. It is important to mention that the GMPE of Boore et al. (2014) has been developed to predict the amplitude-based IMs of the earthquake events, i.e. for finding the IMs such as PGA and SA at a specific vibrational period. This GMPE would be indeed required to be applied to an inverse problem-solving procedure. Since the motion duration of the candidate earthquakes is intended to be simulated using a compatible GMPE on strong motion duration, first we should use the GMPE of Boore et al. (2014) in order to convert an amplitude-based IM to a corresponding moment magnitude $M$ in the aforementioned inverse problem-solving procedure. In fact, finding the corresponding moment magnitude $M$ at each level of IM can make a suitable link between the considered IM and the input parameters of the compatible GMPE of motion duration. In this case, first all contributing or input-related factors of this GMPE (Boore et al. 2014) along with an IM value such as PGA or SA—except for the moment magnitude $M$—are chosen based on the characteristics of the scenario earthquakes expected at a specific site of interest. These input factors, which are fully explained under Section 3.2, are then employed for use in Equation (9) to solve for a corresponding value of $M$ that will be used instead of the selected IM value in the next step. In this way, a specific level of IM is exchanged with an equivalent or corresponding $M$ obtained here in this step.

$$\ln Y = F_E(M, \text{mech}) + F_p(R_{JB}, M, \text{region}) + F_S(V_{S30}, R_{JB}, M, \text{region}, z_1) + \varepsilon_n \sigma(M, R_{JB}, V_{S30})$$ (9)

where $\ln Y$ is the natural log of an amplitude-based IM such as PGA, PGV or 5% damped pseudo spectral acceleration (PSA); $F_E$ and $F_P$ are functions related to the source- and path-term parameters ("events"); $F_S$ is a function for site term. $\varepsilon_n$ is an error term, and $\sigma$ is a function representing the total standard deviation of the model. For further detailed explanation about the functions reflected in the Equation (9), please refer to Boore et al. (2014). Parameter *"region"*, in the above equation, can have different values that are devised to change the region in which this relation is going to be employed. Also, the term *"mech"* denotes fault type mechanism. The rest of the independent variables of each function, such as Vs30 and $R_{Jb}$, used in Equation (9) are explained in Section 3.2.

It is essential to add that since the GMPE of Boore et al. (2014) and the other NGA-West2 models have been derived from the data mainly related to the California region and considering a fact that we are going to limit our numerical examples of this study to the California seismic zone, it can be readily expected that the ground motion parameters pertinent to the California area would be appropriately estimated by these GMPEs without a need for any further validation. Moreover, Gregor et al. (2014) reported that NGA-West2 models used in different areas outside of California—but within the regions these equations are applicable such as China, Japan, Italy—can only have different outcomes when larger distances more than 80 km are going to be considered. However, it is highly recommended to follow a validation procedure against the real recorded data, as suggested by Zalbuin et al. (2018) and Likitlersuang et al. (2020), for the predictions of the GM parameters in those areas where the performance of these GMPEs are still under question.

ii. To have a target acceleration spectrum at each level of IM, one can start with an initial value of IM, e.g. a PGA value or of a *Sa (T=T1, μ=5%)*. First, this initial value of IM is changed with a corresponding or equivalent value of $M$ in the previous step. Then the equivalent moment magnitude $M$ and the other input characteristics of the scenario earthquake, which are expected at a specific site of interest, would be utilized as a new set of input parameters for the GM model (here, Boore et al. 2014) selected to produce the data related to amplitude-based IMs of the potential scenario earthquakes. With these new input parameters in hand, the realization of *Sa (T, μ=5%)* can be generated by the GM model of Boore et al. (2014) at the considered IM level. Finally, the median value of *Sa (T, μ=5%)* from the produced realizations would be taken as the representing acceleration spectrum at this IM level. For having different acceleration spectra at different levels of IM, the procedure described herein in this step should be separately repeated for each level of IMs considered for the analysis.

iii. The motion duration of the scenario earthquakes is subsequently generated for each level of IM, where an MCS-based simulation process would be employed in this proposed framework. In this regard, a compatible GMPE should be considered and selected for motion duration of the earthquake events whose characteristics—including the moment magnitude M—have been nominated from the previous steps. As said earlier while there are many definitions for the motion duration (Bommer and Martinez-Periera 1999), the definition for the significant duration is selected as a duration-related parameter in this paper because it is a continuous time interval on the ground motion accelerations and can be reliably estimated by a compatible ground motion model developed by Afshari and Stewart (2016). This GMPE—which is on significant duration D5-75—is employed for the MCS-based data sampling of this study. As mentioned before, the duration metric selected from the previous section should be consistent with the GMPE one selects for the data sampling of motion duration. So, the compatible GMPE selected for the duration simulation dictate a duration metric that the GM equation is based on. As reflected in Equation (10), the standard error term ($\varepsilon_n$) of this GMPE is considered as a random variable for this data sampling simulation.

$$\ln D = \ln\left(F_{Ed}\left(M, \text{mech}\right)\right) + F_{pd}\left(R_{\text{rup}}\right) + F_{sd}\left(V_{S30}, \delta z_1\right) + \varepsilon_n \sigma_d\left(M\right) \tag{10}$$

where ln$D$ is the natural log of significant duration; $F_{Ed}$ and $F_{Pd}$ are functions derived to stand for the source- and path-term parameters; $F_{Sd}$ is also a function to represent site term; $\sigma_d$ is a function for the total standard deviation of the significant duration model. For additional explanations about the prediction model of Equation (10), readers are referred to see Afshari and Stewart (2016). The independent variables of each function used in Equation (10), are adequately described in Section 3.2.

It is worth mentioning that the GMPE derived by Afshari and Stewart (2016) is quite compatible with the GMPE chosen in the previous steps—the GMPE recommended by Boore et al. (2014)—because both of them have been derived from the same identical number of events and database, namely the NGA-West2, and had the same screening protocol as well. It may be useful to recall that Boore et al. (2014) have been selected in this study for the sampling of the data related to amplitude-based IMs of this proposed joint simulation that is between motion duration and the amplitude-based IMs. Hence, the conditions for the compatibility of these two GMPEs for this joint simulation can be effectively satisfied in this regard.

iv. In a procedure like the one described in step 2 of this section, the realization of motion duration values of the simulated events at each IM level is individually generated by this simulation procedure using the GMPE of strong motion duration. For this purpose, the moment magnitude—exchanged with an initial IM value in step 1—along with the other expected input parameters of the scenario earthquake would be selected as a set of new input factors for the GM model of the motion duration, here the GMPE of Afshari and Stewart (2016). The median value of the generated realizations at each IM level considered would be picked as the median motion duration at that level of IM. If we continue to perform such data sampling for all levels of IM selected to be at work for our proposed method, a relationship function, i.e. *dur=f$_{Dur}$(IM)*, can be obtained through connecting the points related to the median motion duration at each level of IM using either a moving average or a least-square optimization method. Whenever such a function has been developed, the median duration at a specific level of IM can be easily estimated. In fact, median duration values would be continuously and conveniently available for all required intensity levels.

A flowchart for the aforementioned simulation process is provided in Figure 3 where the term "dIM" stands for each increment of IMs employed in the proposed method. As can be seen from Figure 3, the site and fault parameters of a scenario earthquake should be specified before the simulation procedure. These parameters which are controlling the data sampling of a specific site are adequately described in Section 3.2.

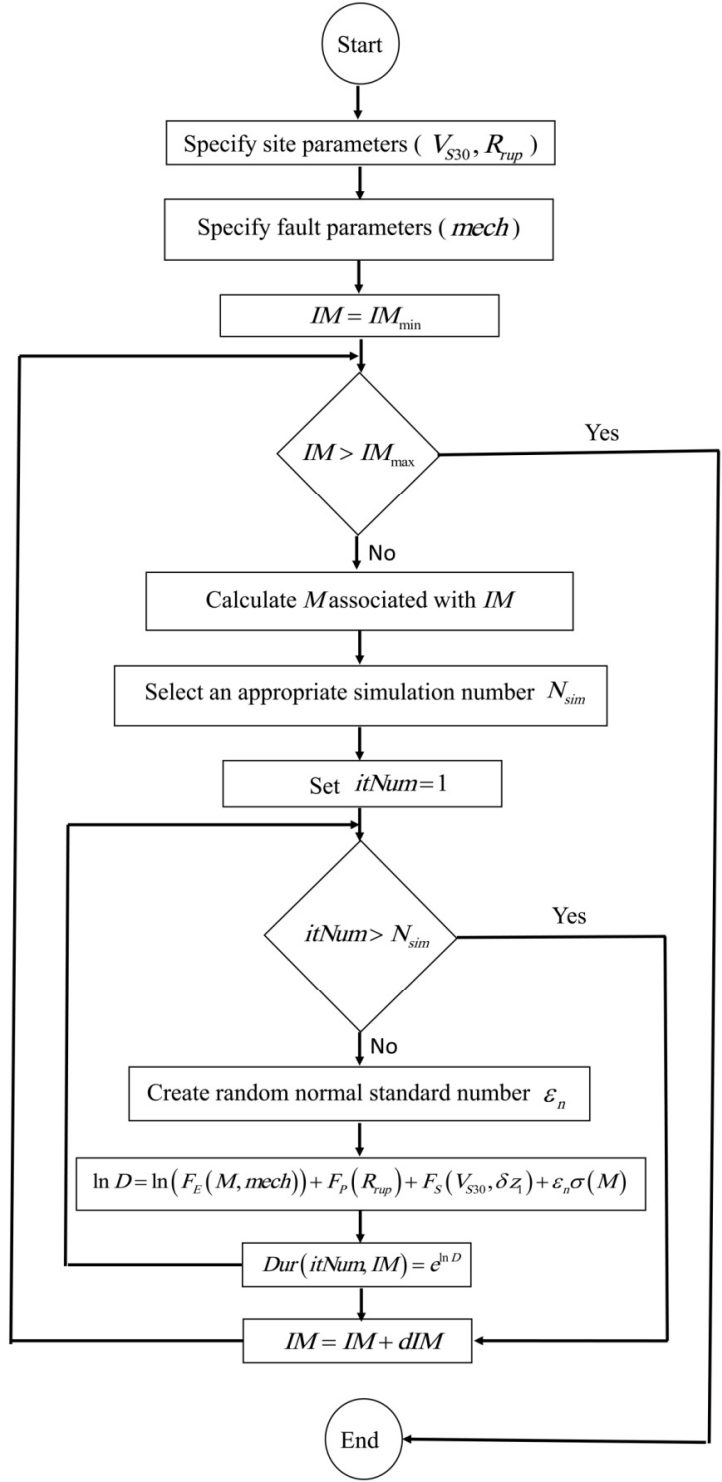

Figure 3. A step-by-step algorithm for the simulation procedure of this study, which is proposed for the incorporation of motion duration in the incremental-based dynamic assessments

As mentioned earlier in the methodology section, with the target acceleration spectrum at each IM level derived in step 2, and the obtained relationship function between the median motion duration and the candidate IM metric from step 4, a set of GMs compatible with these two targets can be readily produced at each level of IM. As stated in the previous section, a set of artificial GMs compatible with the target response spectrum and target duration can be generated at each level of IM considered for the structural analysis of the proposed method. While there are several methods for the generation of artificial spectrum-compatible GMs in the literature (e.g. Saragoni and Hart 1973, Boore (1983, 2003), Atkinson and Silva 2000), in this study, a time-domain approach developed by Gasparini and Vanmarcke (1976) has been selected for the simulation of the required artificial GMs.

Gasparini and Vanmarcke (1976) suggest that first a set of sinusoidal periodic functions—with random phase angles and different amplitudes—be generated as the initial guesses. So with outlining a vector of amplitudes and simulating different arrays of phase angles, it is expected to have different processes with the same general aspect but with different characteristics. It should be noted that these processes are stationary and their properties do not change with time. To impose and create the transient nature of the earthquakes, the steady-state motions (or random seeds) are multiplied by deterministic envelope shape to get a modified random seed. In an iterative manner, the new amplitudes of the modified random seeds are next computed and found using a Power Spectral Density Function (PSDF). This procedure, the calculation of PSDF that is based on the velocity target spectrum, would be continued until then a convergence status is reached. When the difference between the velocity spectrum (or the response spectrum equivalently) of the modified seed and the target is less than a predefined misfit value, the artificially generated GM meets the convergence criterion (Gasparini and Vanmarcke 1976). It is worth mentioning that different envelop shapes, such as an Exponential (Liu 1969) or a Saragoni & Hart (Saragoni and Hart 1973) envelope functions, can be utilized for the generation of the modified random seeds as the initial points. After selection of an envelope shape, the ratio of the total duration of the modified seed to a selected duration-related parameter—namely, the D5-75 metric—can be appropriately approximated. So, many modified random seeds would be generated with this approximated total duration of motion. Next, the ones that can meet the aforementioned convergence criterion would be selected for the next round. In this stage, the D5-75 parameter of all converged GMs would be individually calculated. But only records with a misfit value, between the calculated D5-75 and the target D5-75, less than a predefined threshold (the duration misfit) would be selected as the final GMs at a specific level of IM.

Generally, in this proposed method the concept of spectral matching for the dynamic analysis is applied to the potential artificial or modified GMs at each level of IM. Since there is a large variability in the unmatched GMs, more dynamic analyses may be required to compute the seismic response of the structures because the dispersions of the structural response would be large in such cases. This can be a time-consuming and expensive procedure in case a dynamic analysis for complex structures is under investigation because more numbers of unmatched GMs may be essential to calculate the median response of a structure (Jayaram et al. 2011). Consequently, matched GMs can substantially reduce the computational time related to such dynamic analyses since a lower number of matched records compared to the unmatched GMs would be needed in this case (Bazzurro and Luco 2006, Watson-Lamprey and Abrahamson 2006, Carballo 2000, Shahryari et al. 2019). Accordingly, it is interesting to find out that we can get the benefits of the reduced number of matched motions in the proposed method because both artificial and modified GMs are adequately matched to a target response spectrum at each level of IM.

As mentioned earlier, instead of the production of artificial earthquake motions, several real GMs compatible with the target response spectrum at each level of IM can be also selected as the initial guesses for the GMs at an IM level. But it should be noted that the GMs selected should have a duration misfit value less than the one defined for target duration at each level of IM. Next, a spectral matching procedure—like the one proposed by Hancock et al. (2006)—would be applied to the duration-compatible GMs selected up to here. So, with appropriate misfit values for the SA as suggested in Equation (6), the selected GMs would be adjusted in a way that their acceleration spectra get matched to a target response spectrum that is defined by the simulation procedure at each level of IM. To end with, the modified GMs having a misfit value less than a predefined threshold would be picked as the final GMs for an IM level. For more details related to the spectral matching procedures, the readers are invited to make a review on the available literature for this subject (Hancock et al. 2006, Atik and Abrahamson 2010).

## 3.2. Parameters of the earthquake scenario

In order to determine a relationship function between motion duration and a candidate IM, variables pertinent to the simulation process of a given site should be first identified. These variables are normally the ones appear in the GMPEs which are both derived for motion duration and amplitude-based IM of the earthquake events. In the following paragraphs, variables associated with source, sites and sour-to-site distances are adequately described.

Soil condition parameter is expressed by means of the time-averaged shear wave velocity over a sub-surface depth 30 meters, denoted as Vs30. For site classification regarding Vs30 parameter, this study uses the recommendations provided by NEHRP provisions (1997). Thus, first we need to find depth parameters—the so-called $Z_{1.0}$ and $Z_{2.5}$—in order to fully characterize the Vs30 term in the simulation procedure. Depth parameters $Z_{1.0}$ and $Z_{2.5}$ are defined as the depth level at which shear wave velocity reach 1000 m/s and 2500 m/s, respectively. The $Z_{1.0}$ depends on the Vs30 and is calculated according to a relationship developed by Abrahamson and Silva (2008) as expressed by Equation (11). The $Z_{2.5}$ is then computed by an extrapolation procedure based on $Z_{1.0}$ parameter as recommended by Campbell and Bozorgnia (2006).

$$Z_{1.0} = \begin{cases} \exp(6.745) & V_{S30} < 180 m/s \\ \exp\left[6.745 - 1.35\ln\left(\frac{V_{S30}}{180}\right)\right] & 180 \leq V_{S30} \leq 500 m/s \\ \exp\left[5.394 - 4.48\ln\left(\frac{V_{S30}}{500}\right)\right] & V_{S30} > 500 m/s \end{cases} \quad (11)$$

Multitude distance measures are recommended to describe the distance-related parameters, including the rupture and Joyner-Boore distance. The rupture distance ($R_{rup}$)—which is defined as the slant distance to the closest point on the rupture plane—is employed to be as a constant variable hereafter because there is regularly a constant distance between the faulting point and where ground motions are recorded. In this case, while earthquakes are always originated from a single seismic source, several different distances (i.e. 10, 15 and 20 km) from an active fault have been chosen for this investigation. This can cover a range of distances from near-source events to the ones related to the far-sources. In addition to rupture distance, existing GMPEs may also need the Joyner-Boore distance ($R_{JB}$) which is defined as a horizontal distance to the surface projection of the rupture. In this study the Joyner-Boore distance is assumed to be equal to the rupture distance for the sake of simplicity.

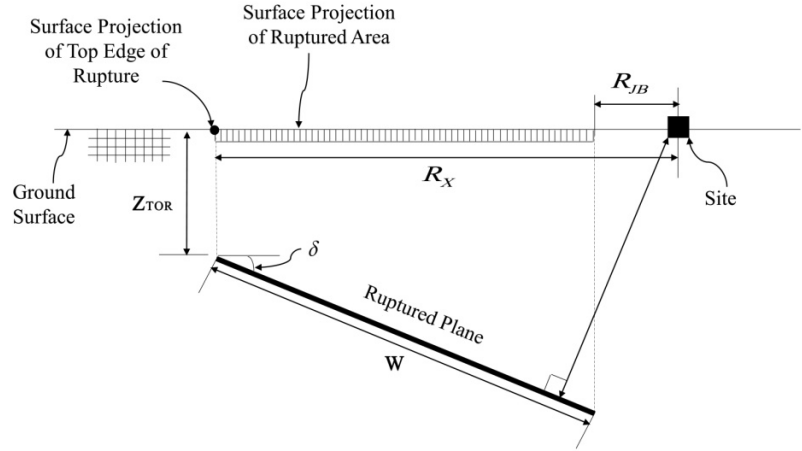

Figure 4. Schematic illustration of earthquake source and distance measures using a vertical cross section through fault rupture plane (Kaklamanos et al. 2011)

In Figure 4, $\delta$ is fault dip, $W$ is down-dip rupture width, and $Z_{TOR}$ is depth-to-top of the rupture. *Dip* (or $\delta$) is the angle that a planar geologic surface is inclined from the horizontal one, where strike-slip faults are assumed to be vertical ($\delta=90$) in this case. Moreover, the average values of dip angle equal to 50 and 40 degrees are recommended for normal and reverse faulting events, respectively (Kaklamanos et al. 2011). $R_X$ is the horizontal distance to the surface projection of the top edge of the rupture, which is measured perpendicular to the fault strike and is computed by Equation (12) as suggested by Kaklamanos et al. (2011):

$$R_X = \begin{cases} R_{JB} |\tan \alpha| & 0 \leq \alpha < 90 \text{ and } 90 < \alpha \leq 180, \ R_{JB}|\tan \alpha| \leq W \cos \delta \\ R_{JB} \tan \alpha \cos\left[\alpha - \sin^{-1}\left(\dfrac{W \cos \delta \cos \alpha}{R_{JB}}\right)\right] & 0 \leq \alpha < 90 \text{ and } 90 < \alpha \leq 180, \ R_{JB}|\tan \alpha| > W \cos \delta \\ R_{JB} + W \cos \delta & \alpha = 90, R_{JB} > 0 \\ \sqrt{R_{RUP}^2 - Z_{TOR}^2} & \alpha = 90, R_{JB} = 0, R_{RUP} < Z_{TOR} \sec \delta \\ R_{RUP} \csc \delta - Z_{TOR} \cot \delta & \alpha = 90, R_{JB} = 0, R_{RUP} \geq Z_{TOR} \sec \delta \\ R_{JB} \sin \alpha & -180 \leq \alpha < 0 \end{cases} \quad (12)$$

where α is the source to site azimuth that for a given site is the angle between the positive fault strike direction and the line connecting the site to the closest point on the surface projection of the tope edge of the rupture (Chiou 2005). This angle is assumed positive when it is measured clockwise as shown in Figure 5.

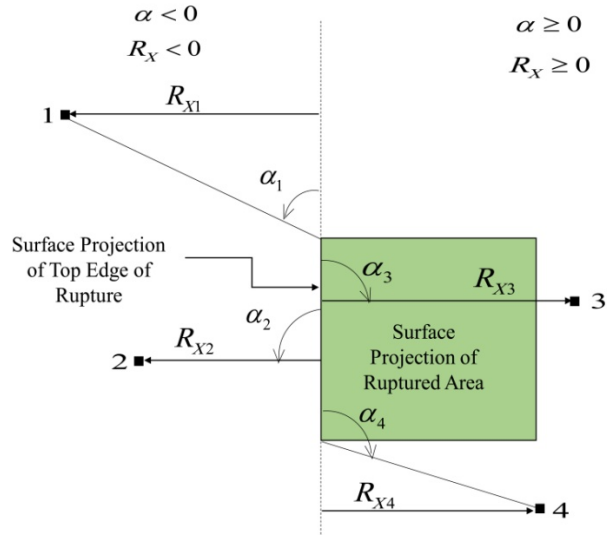

Figure 5. Plan view of a fault rupture (Kaklamanos et al. 2011)

For computing the distance-related parameter $R_X$ from Equation (12), the relationship developed by Wells and Coppersmith (1994) is used to estimate the down-dip rupture width (W) from moment magnitude and the style of faulting as brought in Equation (13). It is worthy to add that different fault types can be considered for simulation procedure—namely the normal, reverse and strike-slip.

$$W = \begin{cases} 10^{-0.76+0.27M} & \text{for strike-slip events} \\ 10^{-1.61+0.41M} & \text{for reverse events} \\ 10^{-1.14+0.35M} & \text{for normal events} \end{cases} \quad (13)$$

As can be readily understood, a term standing for the depth-to-top-of-rupture ($Z_{TOR}$) should be determined as an input variable of Equation (12). The method employed by Kaklamanos et al. (1994) is also used here to estimate $Z_{TOR}$ from hypocentral depth ($Z_{HYP}$), down-dip rupture width (W), and dip angle ($\delta$) as expressed in Equation (14):

$$Z_{TOR} = \max\left[(Z_{HYP} - 0.6W \sin \delta), 0\right] \tag{14}$$

where $Z_{HYP}$ is hypocentral depth, which can be determined according to Equation (15):

$$Z_{HYP} = \begin{cases} 5.63 + 0.68M & \text{for strike-slip faulting} \\ 11.24 - 0.2M & \text{for non-strike slip faulting} \\ 7.08 + 0.61M & \text{for general (unspecified) faulting} \end{cases} \tag{15}$$

### 3.3. Validation of the predicted motion duration against the real data

As can be readily understood from previous sections, thousands of possible earthquake scenarios can be simulated by the proposed data sampling, which seems to be in contrast to the use of real GMs which are limited to a finite number of motions recorded from former events. Hence, a potential advantage of using simulated data is that it might be possible to seek the relationship of a duration-related parameter and an amplitude-based IM with adequate amounts of data, especially for the higher levels of IM. Nevertheless, to evaluate the authenticity and robustness of the proposed MCS-based simulation, characteristics of four real earthquake records have been randomly selected from the NGA-West2 in this regard. These GM records are represented by their Record Sequence Number (RSN) in the NGA-West2 database. For example, a ground motion record with RSN=200 is one of the records obtained from the Imperial Valley earthquake happened in 1979. The essential information related to the main characteristics of the selected earthquakes is presented in Table 1.

Table 1. Essential characteristics of the earthquake records randomly selected for validation procedure

| RSN | Earthquake Name (year) | Station | Magnitude | RJB (km) | Vs30 (m/s) |
|---|---|---|---|---|---|
| 6151 | Tottori, Japan (2000) | EHMH04 | 6.61 Mw | 141.16 | 254.35 |
| 200 | Imperial Valley-07 (1979) | El Centro Array #11 | 5.01 Mw | 13.61 | 196.25 |
| 5151 | Chuetsu-oki (2007) | KNG002 | 6.80 Mw | 227.32 | 113.57 |
| 10000 | Pakistan earthquake (2005) | Thermal Airport | 4.59 ML | 314.99 | 217.8 |

The real characteristics of these GMs are first chosen as the input parameters for step 3 of the proposed simulation procedure presented in Section 2. Then, thousands of motion duration are simulated using the input characteristics of each GM record. Next, a normalized probability density function is obtained based on the simulated data provided by the characteristics of each earthquake event. Figure 6 demonstrates a comparison between simulated data for motion duration of the scenario earthquakes and the ones directly recorded for each of the scenario earthquakes we randomly selected from the NGA-West2 database. In this figure, the vertical blue lines stand for the value of real motion duration of the selected earthquakes, and the curves in red color contain thousands of simulated data for motion duration of the scenario earthquakes. As can be seen in Figure 6 (a) to (d), the median value of the simulated motion duration samples in red curves can well estimate the values representing the real motion duration (being read by blue lines) of the picked GMs.

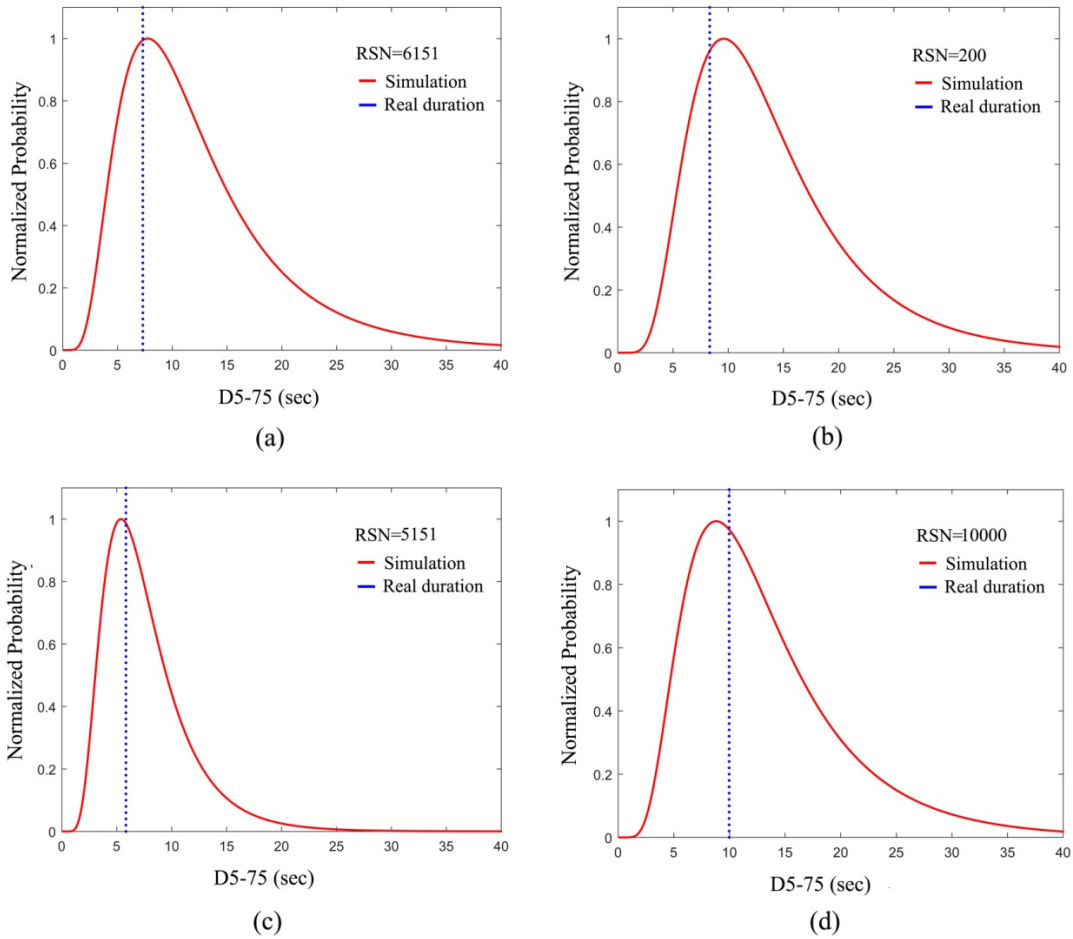

Figure 6. Simulated samples of motion duration versus the values representing the corresponding quantities of real ground motions: a) 2000 Tottori earthquake; b) 1979 Imperial valley earthquake; c) 2007 Chuetsu earthquake; d) 2005 Pakistan earthquake

### 3.4. Relationship functions between duration and PGA

To show the pattern and trend of the simulated data for the variability of motion duration at each level of IM and getting an insight on this matter, a selected site located in California with the parameters described below are employed in this section for developing the relationship function between PGA and the significant duration or 5-75% of normalized Arias-Intensity.

- Vs30= 400 m/s
- Fault type= Reverse
- $R_{rup}$= 15 km

Number of simulation iteration equal to 5000 is used for this considered site. PGA is used as an IM metric and is explored in the interval of [0.001g-0.4g] with an increment equal to 0.001g. In this case, 1,955,000 numbers of simulations should be carried out to derive the relationship function for the site of interest, where the mentioned numbers of simulations have been calculated through the multiplication of the numbers of sampling for the variables involved in this simulation procedure. In other words, 399 IM levels are considered and 5000 simulations are performed at each level individually, so the total number of simulations would be equal to 1,955,000 as mentioned earlier. Raw data points generated through this simulation procedure are depicted in Figure 7. While insufficient real data for large-magnitude earthquakes can cause computational difficulties for statistical analyses—especially for higher IM levels—to assess a significant characteristic of the scenario earthquakes, simulation procedure is performed to such an extent that the number of events at different IM levels is nearly the same as can be seen in Figure 7. This is because the confidence interval length has an inverse relation with the number of samples (Musson 2000), so providing an equal number of events for all possible IMs is the potential advantage of this method. As can be seen in this figure, scattering of the raw data points is much more pronounced at the higher level of IMs; however, the increasing trend of the simulated events is apparently recognizable. It is noteworthy to mention that the observed scattering is mainly due to the dispersion of the data obtained by the selected GMPE.

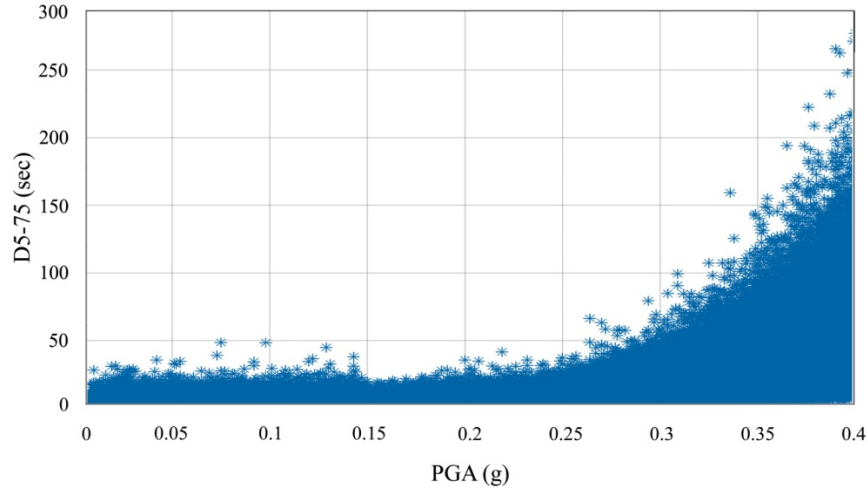

Figure 7. Raw simulated data results for the site of interest (1,955,000 data points)

An exponential function form as brought in Equation (16) is fitted to the generated data. This fitted function along with the median of the generated data points are shown in Figure 8. In Equation (16), a, b, c, and d parameters are constant values which can be readily determined by a simple least-square optimization approach.

$$f(PGA) = a + be^{c(PGA-d)} \tag{16}$$

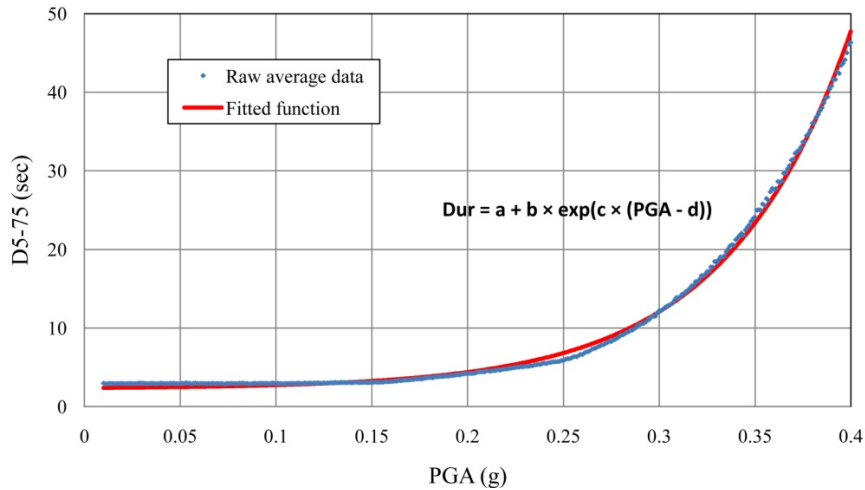

Figure 8. The relationship function between PGA and D5-75 parameter

The influence of different factors on the relationship function between IM and motion duration is also investigated in this section. In this case, the rupture distance and soil condition are considered for the sensitivity analysis. Figure 9 compares the influence of different rupture distances on the relationship functions derived for a specific site in California region. An exponential function is fitted to the data points derived for each distinct

simulation—cases with 10, 15 and 20 km as shown in Figure 9. As it is readily evident, a region with a steady growth rate can be found at the beginning of each curve. It should be added that the faulting mechanism and Vs30 parameter in this example are considered to be equal to the characteristics of the scenario earthquake we introduced at the beginning of this section.

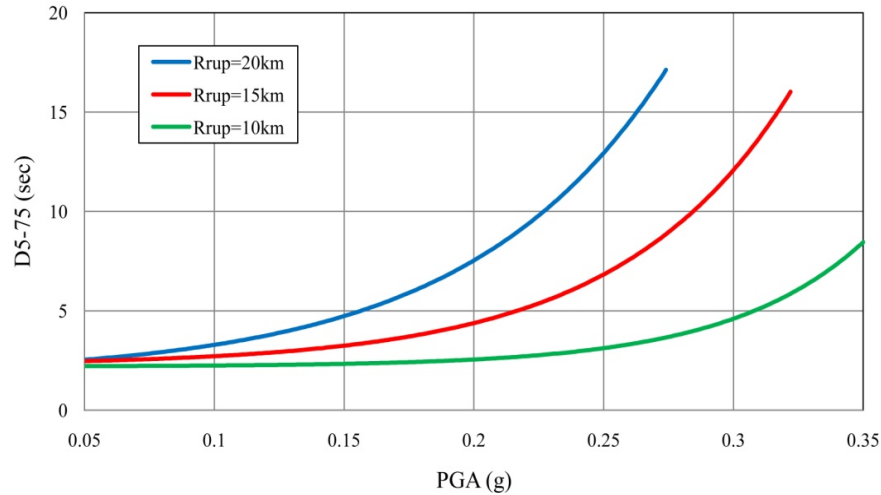

Figure 9. The relationship functions between motion duration and PGA as an IM for three different rupture distances

Figure 9 that includes three rupture distances, i.e. a 10 km, 15 km, and 20 km, shows that if larger values of rupture distance are selected, more variations can be seen for the duration parameter in the relationship functions. As can be seen from Figure 9, larger values for the motion duration at a given level of IM are predicted by the simulation procedure whenever farther source-to-site distances are selected to be employed in this case. It is mainly due to the fact that at farther source-to-site distances (rupture distance=20 km for example), a higher value of moment magnitude—which corresponds to a larger median motion duration—is required to cause an earthquake with larger PGA value. For making this matter much clearer, Figure 9 has been broken into its two components using the moment magnitude M as the intermediary variable. In Figure 10, two components of Figure 9 are created based on the simulation characteristics we used to create the relationships between duration and PGA for different source-to-site distances. In this case, each of the involved variables—motion duration or PGA—is separately drawn against an identical variable used as the intermediary variable, the moment magnitude. As can be seen from Figure 10 (a), for each value of moment magnitude, there is a reciprocal relationship between PGA and the source-to-site distance of the events used for this simulation. This reciprocal relationship between PGA and distance-related parameters of the earthquakes have been reported by several studies (Wang et al. 2002, Shoji et al. 2005). For the other component of Figure 9, relationships of motion duration and moment magnitude (M) as the intermediary variable are generated in Figure 10 (b). As it is shown in this figure, the simulation procedure used to create a relationship between motion duration and moment magnitude as the intermediary variable is not so sensitive to the source-to-site distance. This insensitivity behavior seems to be related to the values chosen as the rupture distances, i.e. a 10 km, 15 km, and 20 km, which are quite near to each other.

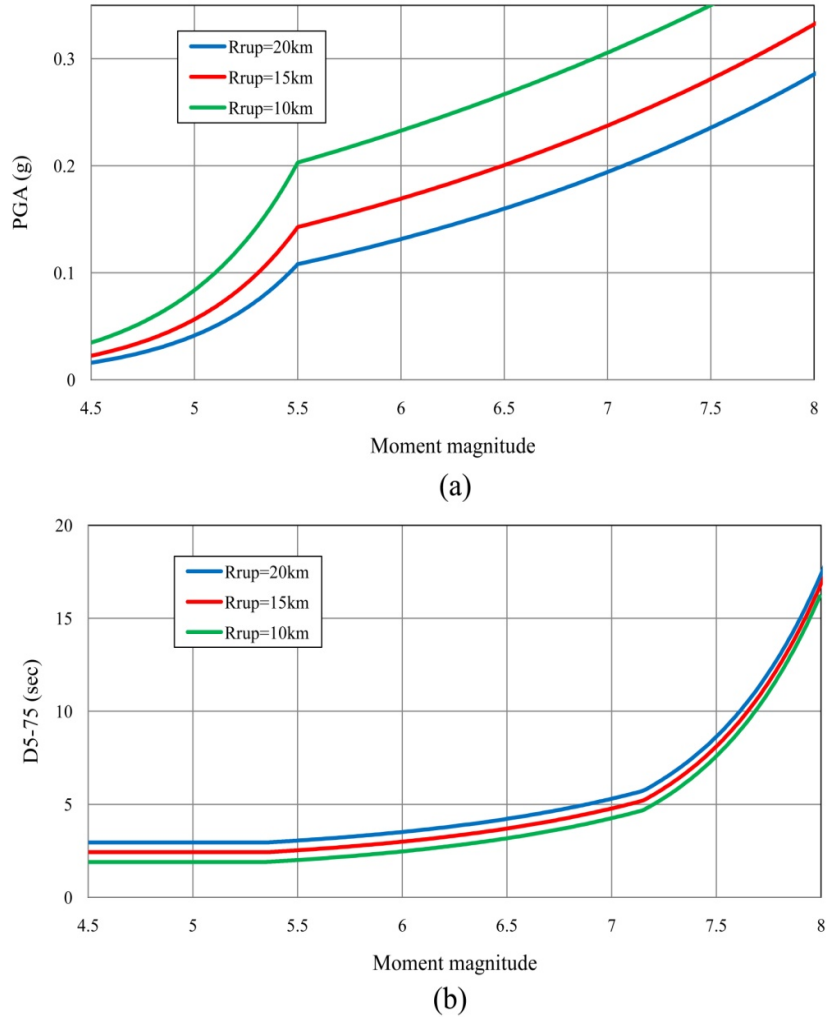

Figure 10. The relationship functions of moment magnitude against the earthquake IMs for three different rupture distances: a) moment magnitude versus the PGA; b) moment magnitude versus the D5-75 parameter

Figure 11 compares the influence of Vs30 values on the derived relationship function, where three types of Vs30 are considered—namely 400m/s, 600m/s, and 1000m/s. As it is quite obvious, a section with a steady growth rate can be also found at the onset of these fitted exponential functions of PGA against D5-75, which gets a bit expanded in case smaller Vs30 parameters are taken to be applied. It should be added that the faulting mechanism and rupture distance in this example are considered to be equal to the characteristics of the scenario event we presented at the beginning of this section.

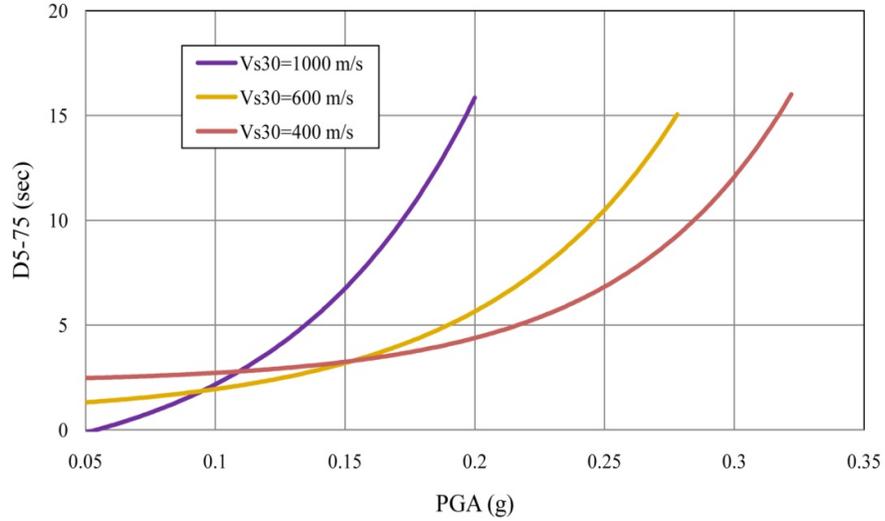

Figure 11. The relationship function between PGA and D5-75 for a selected site with a $R_{rup}$=15 km and reverse fault type but for three different Vs30 values

The relationship curves depicted in Figure 11 reveal that the median predicted values of motion duration at a given level of IM, here for a specific value of PGA, are much higher in case earthquake events are simulated for sites with higher values of Vs30. To find a clue on this matter, each of the curves in Figure 11 is decomposed into two distinct components as it is demonstrated in Figure 12. In fact, each of the involved variables in Figure 11—PGA and motion duration—is drawn against the moment magnitude which is considered as an intermediary variable for the proposed simulation procedure. As can be seen from Figure 12 (a), for each considered value of moment magnitude, there is a reciprocal relationship between PGA and Vs30 parameter. This matter that was reported by several researchers demonstrates that a larger value of moment magnitude is required for stiffer soils to generate the same value of PGA occurring at sites with smaller Vs30 factor (Shoji et al. 2005). Therefore, this reciprocal behavior of PGA and Vs30 along with the information provided in Figure 12 (b) can show us that at a specific level of IM or PGA for stiffer soils, there is a need for a larger value of moment magnitude that is associated with a higher value of median motion duration.

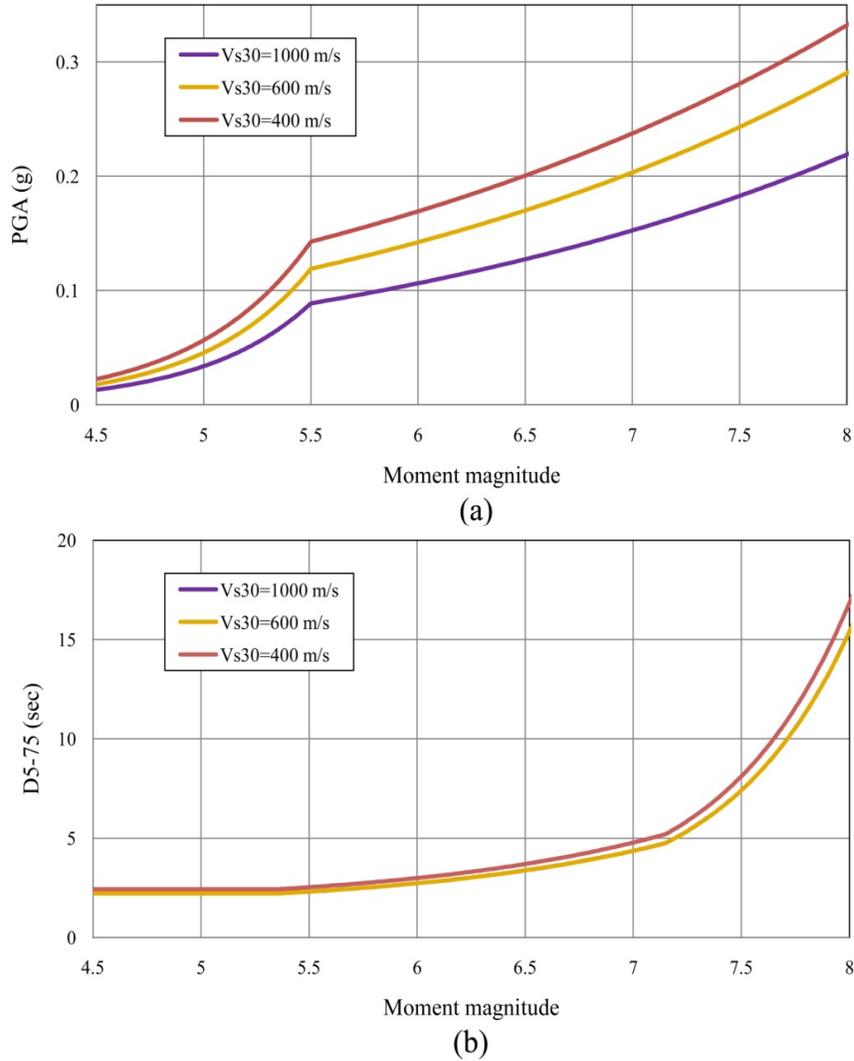

Figure 12. The relationship functions of moment magnitude against the earthquake IMs for a site with a $R_{rup}$=15 km and reverse fault type, but for three different Vs30 values: a) moment magnitude versus the PGA; b) moment magnitude versus the D5-75 parameter

## 4. Application of the proposed method in nonlinear seismic assessments: A case study
### 4.1. Duration-consistent response spectra as a function of intensity measure

For performing a duration-consistent IDA procedure, we need to have a set of duration-consistent GMs for each IM level considered. But we also need a number of duration-consistent response spectra for each level of considered IM before we go ahead with the simulation of the aforementioned duration-consistent GMs. To clarify this matter, an example is taken for a specific scenario earthquake. For each simulated scenario earthquake with $R_{rup}$ = 20 km, Vs30 = 400 m/s and reverse fault type, first an acceleration response spectrum and its related duration parameter are generated. Based on the proposed method of this paper, the median values of the spectral acceleration for a range of first-mode periods and motion duration are then computed for the simulated data of each sample or each IM level. In order to produce duration-consistent GMs, the median acceleration spectrum is generated for 64 levels of IM or in 64 different samples, where each acceleration spectrum is associated with a median D5-75 parameter. In other words, these motions—which can then be used for the application of nonlinear seismic assessment—are based on the median duration-sensitive spectra that are produced in 64 levels of SA. Four of these simulated response spectra, with their IM levels and associated duration parameters (D5-75), are displayed in Figure 13. As can be seen from this figure, each spectrum with a specific level of IM is accompanied by a unique D5-75 parameter. Overall, these acceleration spectra along with their corresponding duration parameters are found according to the proposed simulation method. Besides, it is evident to see that spectral shape at each level of IM gets dynamically altered as indicated in Section 2.

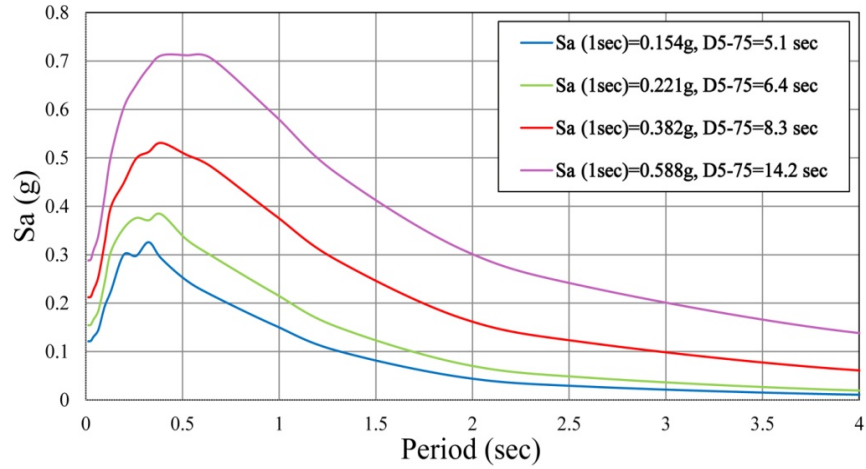

Figure 13. Generated median response spectra with different IMs and D5-75 parameters

### 4.2. Duration-consistent ground motions

Based on the duration-consistent response spectra that can be normally generated from previous step or from Section 4.1, duration-consistent GMs are produced using techniques available in the literature to generate a set of spectrum-compatible artificial GMs. In this study, a set of GMs is simulated for each IM level employing duration-consistent response spectra of the scenario earthquake computed in Section 4.1. Using SeismoArtif (2018) software, a bin of 8 ground motions is generated based on each median spectrum acceleration as well as the associated median D5-75 parameter, where each artificial ground motion is simulated utilizing power spectral density function of Gasparini and Vanmarcke (1976) and the shape function of Saragoni and Hart (1973) as envelope shapes. The response spectra pertinent to a bin of generated motions, with a target spectrum created based on IM=0.382g and D5-75=8.3sec, are plotted in Figure 14. In this figure, the average misfit between the target spectrum and the acceleration spectra of the simulated GMs ranges between 3 to 7 percent, which is all calculated based on the Equation (6) presented in Section 3. Besides, we managed to maintain the duration misfits of the simulated GMs between 2 to 6 percent. It should be also mentioned that more exact methods for the generation of target-based GMs—artificial or spectrally matched motions—can be employed to improve or diminish the average misfit values of the GMs simulated at each level of IM.

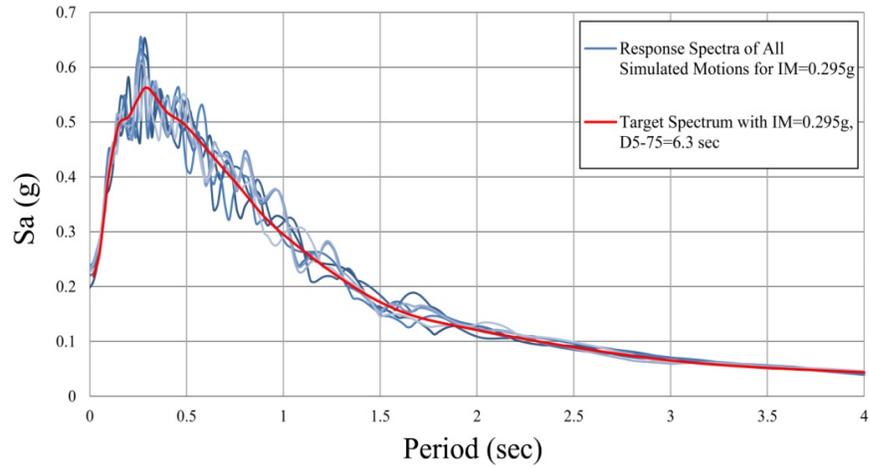

Figure 14. Response spectra of the duration consistent records along with their target spectrum for IM=0.382g and D5-75=8.3sec

Note that duration parameter (D5-75) is generally increased as the SA is increased up to the level of 64. Therefore, 512 duration-consistent GMs are produced, which are subsequently used for the nonlinear seismic assessment. Simulated GMs at four intensity levels, with IMs equal to 0.154g up to 0.588g, are presented in Figure 15 (a) to (d).

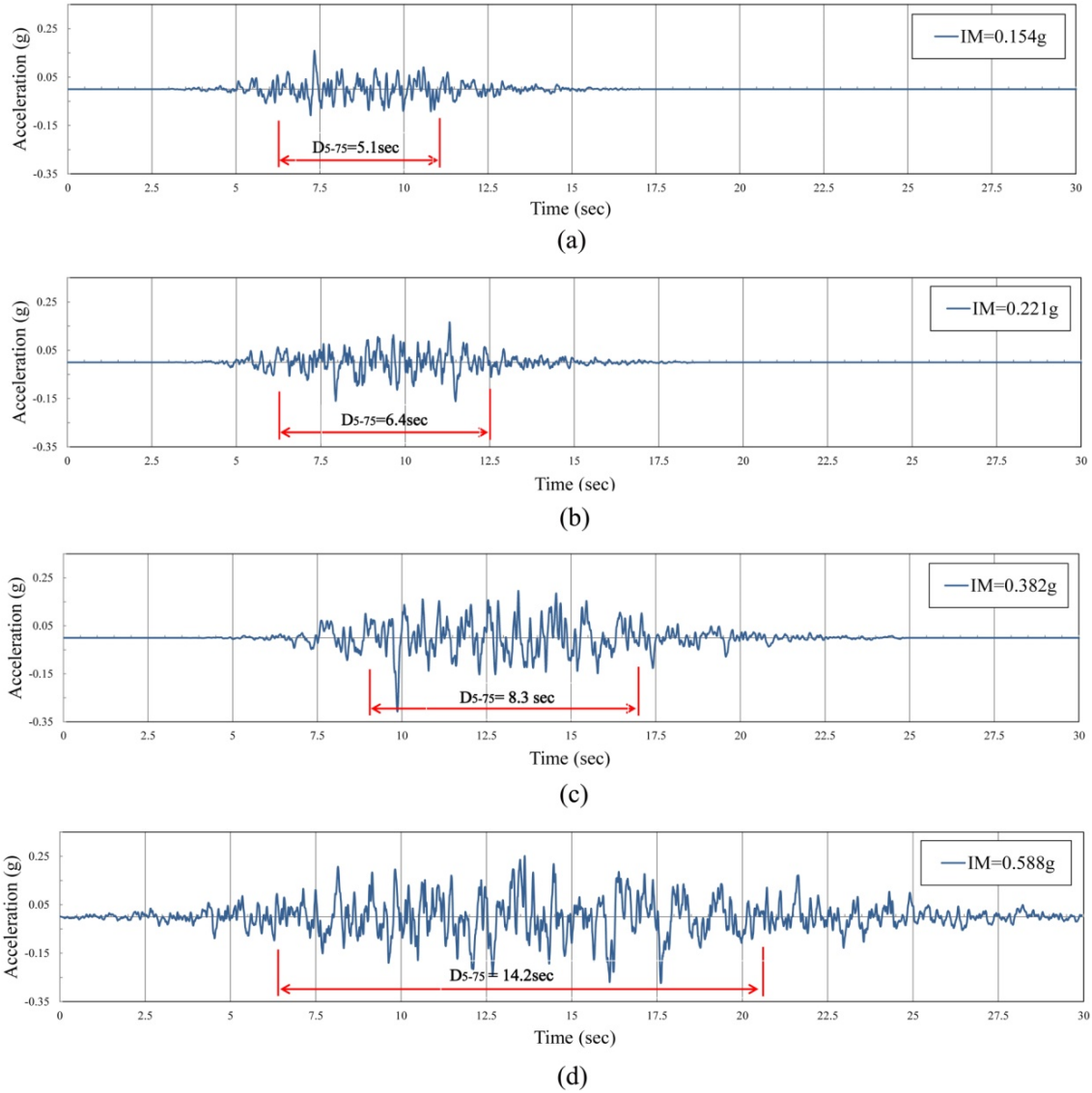

Figure 15. Simulated earthquakes with different IMs and motion durations: a) IM=0.154g, D5-75=5.1sec; b) IM=0.221g, D5-75=6.4sec; c) IM=0.382g, D5-75=8.3sec; d) IM=0.588g, D5-75=14.2sec

### 4.3. Nonlinear seismic assessments of case study models

Two SDOF systems, one with degrading behavior and the other one without degradation, are taken to investigate the application of the proposed method in the nonlinear seismic analysis. In this case, structural SDOF systems with a 1-sec period of vibration are considered to be exposed to a set of duration-consistent GMs generated for the scenario earthquake of Section 4.2. The inelastic SDOF systems with degrading and non-degrading behavior are modeled using bilinear and Ibarra-Krawinkler hysteretic model (Ibarra et al. 2005), respectively. The employed hysteretic models for both degrading and non-degrading SDOF systems are displayed in Figure 16. For both degrading and non-degrading models, $F_y/W$ is set to be 0.082, where $W$ is the weight of SDOF systems and $F_y$ stands for the yield strength. Factors associated with the hardening as well as post-capping stiffness, $a_s$ and $a_c$, are selected to be 0.006 and -0.02, respectievely. Residual strength is also 0.01 of the yield strength ($\lambda$=0.01).

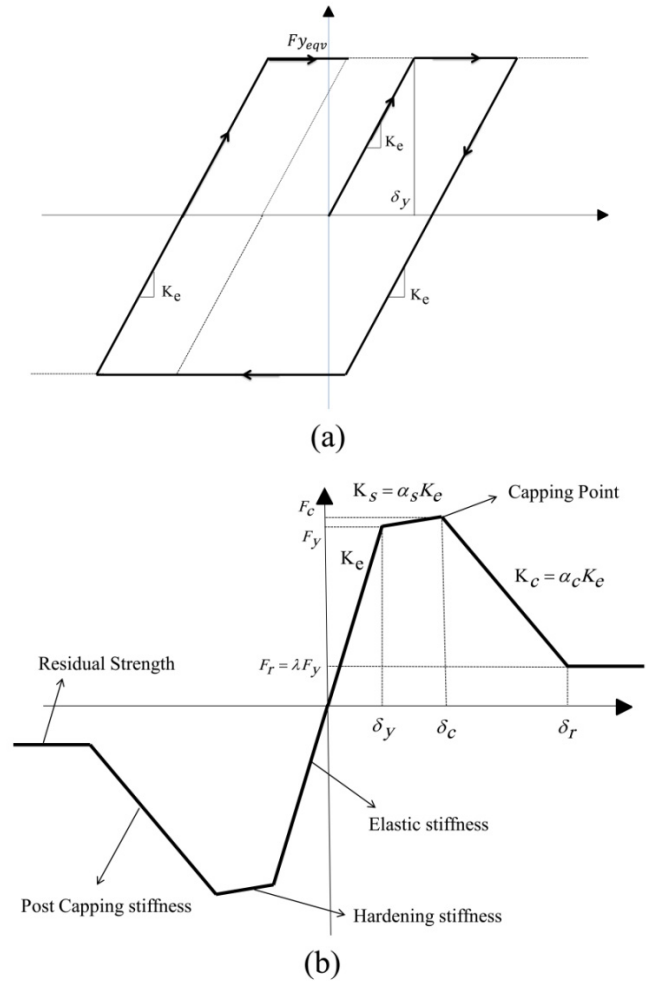

Figure 16. Employed hysteretic models: a) bilinear elastoplastic model utilized for non-degrading SDOF system; b) backbone curve of the Ibarra-Krawinkler used for degrading SDOF system (Ibarra et al. 2005)

The median response curves to predict the seismic behavior of the systems are obtained by two different approaches. First, a procedure similar to the cloud analysis (Mackie and Stojadinovic 2005) is done considering all duration compatible ground motions of this study. This means that a number of nonlinear time history analysis—required at each seismic intensity level—is performed using a bin of ground motions that are compatible with the considered seismic intensity level. In this case, the median response curve is obtained by computing the median value of responses at each level of SA. A standard IDA procedure, which is completely similar to the classical IDA analysis, is also considered afterward. In this case, just one set of ground motions is selected and then scaled for all intensity levels. For this purpose, a bin of ground motions generated for the thirtieth level of SA (with IM=0.382g) is selected as the record set of the IDA procedure. Therefore, scaling down and up procedures should be followed for computing the median response curve in the standard (conventional) IDA framework.

Figure 17 demonstrates the median response curves for the two mentioned seismic response procedures (i.e. the IDA and cloud-like response procedures). These response curves are obtained using two inelastic SDOF systems, with and without degradation capturing capability. Mean ductility factors, $\mu = \delta_{max}/\delta_y$ several levels of seismic intensity are also provided for each of the median response curves presented in Figure 17 (a) and (b). In this case, the term $\delta_{max}$ is the maximum absolute value of displacement response, and $\delta_y$ stands for the yield displacement of the system. As can be seen from this figure, the median response curves are generally in a good agreement for all levels of SA up to the 0.4g while they get widely separated from each other at the higher levels of seismic intensity. It is apparently recognizable that the median response curves computed with cloud-like dynamic analysis, for both models with and without degradation, demonstrate softening initiation around the SA level equal to 0.4g. But the median response curves of the IDA analysis—which is performed with the GM records of IM=0.382g—do not show the aforementioned softening initiation points in such an extent. Besides, the proposed method shows relatively higher ductility factors compared to the standard IDA procedure at the upper levels of seismic intensity, which demonstrates that the considered structural systems in this analytical framework have experienced further nonlinearity. As it is expected and seen in Figure 17, higher values of ductility factor bring more scattering in the computed displacement responses. Therefore, it can be seen that the expected variability of synchronization of motion duration and the amplitude-based characteristics of the ground motions is well incorporated in the proposed incremental dynamic procedure. Needless to say that the mentioned synchronization of motion duration and amplitude-based IM of the records are being reflected through the target response spectrum and the estimated motion duration at each level of IM.

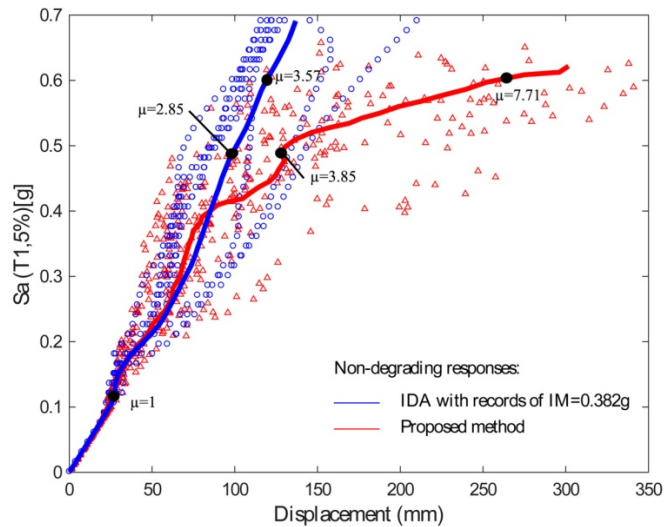

(a)

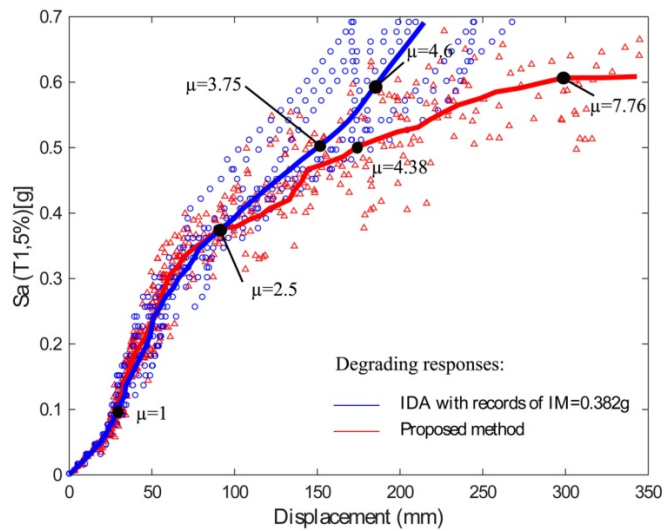

(b)

Figure 17. Median response curves and the associated data related to the response of each individual ground motion: a) for a non-degrading SDOF model with T1=1 sec; b) for a SDOF model with degradation and T1=1 sec

To evaluate and separate the variability of the proposed incremental seismic response in terms of the duration inclusion of motions and the changing nature of spectral shape at each level of IM, another two sets of artificial GMs have been created based on the characteristics of the 512 records simulated for the proposed method—those GMs generated in section 4.2. For the first group of GMs, all records would have the same length of motion duration which is completely equal to the duration length of the GMs chosen for the standard or conventional IDA procedure just performed—i.e. the IDA procedure with the GM records of IM=0.382g. This group of motions is produced according to the different spectral shapes for the IM levels used for the analysis, meaning that the artificial records of this group are not generated with a specific constant target response spectrum for all IM levels. The function of spectral shape at each level of IM is extracted from the simulation procedure, which is the same as the ones used for the generation of artificial GMs just applied to the proposed dynamic framework. So, we can name this group of GMs as a set of records with varied response spectra. For the record set with varied motion durations, new artificial GMs are then simulated based on the same response spectrum which is the mother response acceleration spectrum of the GMs used in the conducted standard IDA. In this case, the records with varied motion durations but constant response spectrum can be regarded as the second group of earthquake records considered for comparative study. These two new groups of GMs—motions with varied durations and varied spectra—along with the earthquake motions generated in section 4.2 would be used to be applied to the degrading SDOF model introduced at the beginning of this section. The results of these incremental dynamic analyses are depicted in Figure 18 (a) to (d). In fact, the outputs of the analyses under three aforementioned groups of GMs are displayed against the median IDA curve computed from the standard IDA procedure with the GM records of IM=0.382g.

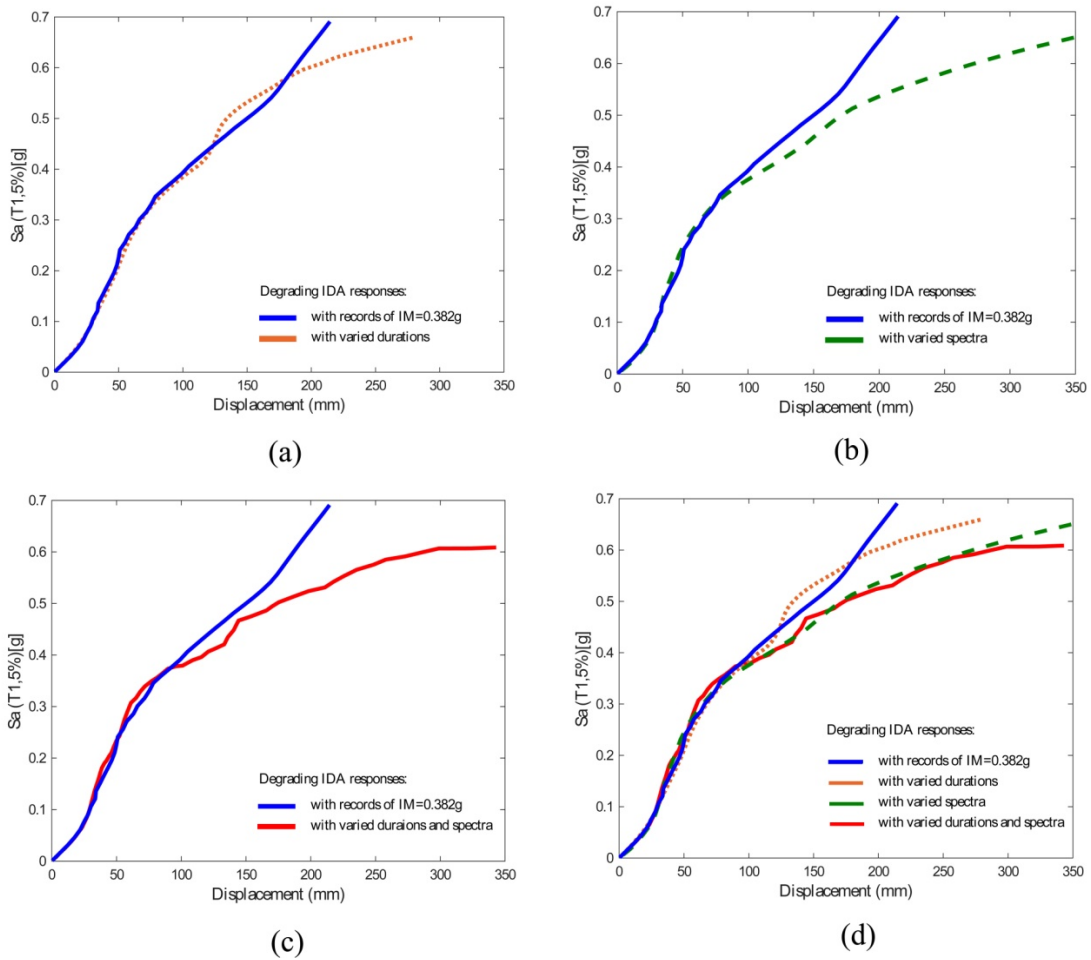

Figure 18. Median IDA response curve against: a) incremental response with varied durations; b) response with varied spectra; c) response of the proposed method; d) all considered responses

In Figure 18 (a), the IDA response curve with records of IM=0.382g is compared with the outputs of dynamic analysis under GMs with varied durations. It should be re-mentioned that the two sets of records used here have the same spectral shapes for all IM levels. As can be seen in this figure, taking into account the variability of motion duration can cause a softening initiation in the IDA curve under varied durations at the higher levels of IM. The same trend and pattern can be seen in Figure 18 (b) which is created to compare the result of standard IDA procedure with the response curve computed using motions with varied response spectra. But when both the variability of duration and response spectra are taken into account at the same time, which is a matter that can be expected from the proposed method, then the result would be different once it is compared to a case considering the variability of duration or the spectral shape in a separate way. As can be seen in Figure 18 (c), it is apparently recognizable that the synchronization of duration and amplitude-based IM of the records can cause the related response curve to get heavily softened and eventually flattened at the higher level of IMs. This matter is more highlighted in Figure 19 which is drawn to show differences between standard median IDA and considered incremental response curves. When the error value of a specific response curve is equal to 0.5 at a given level IM, it means that the value of its displacement response is more than 50% of its counterpart in standard IDA procedure. For the proposed method, the curve representing its error values goes beyond any other error curves, approaching the infinite slope before an SA of 0.6g. The infinite slope of the error curve exhibits signs of dynamic instability or the occurrence of a collapsing mechanism at the higher seismic intensity levels. Actually, it can be understood that the differences witnessed between the median response curves of IDA and the proposed method for the higher level of seismic intensity are due to the fact that the synchronization effect of duration and SA is not suitably incorporated in the standard IDA procedure. The reason may be due to the fact that just the amplitude-based IM (here SA) is scaled in the such conventional IDA procedure while both the level of SA and the related duration of motions are simultaneously changed in the proposed cloud-like response procedure.

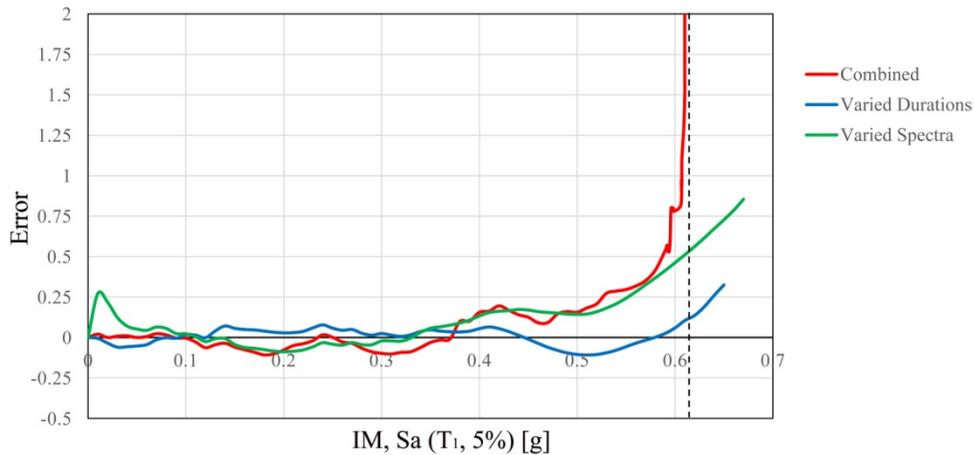

Figure 19. Differences and residues of standard IDA response curves and the rest of considered incremental responses at deferent levels of IM

## 5. Discussion

In the IDA framework, the levels associated with IM are changed by scaling up and down of a set of GMs selected before the analysis. In this case, IMs are monotonically increased from a low level of excitation up to where recognized as a seismic level believed to cause a collapsing mechanism for the structure of interest. In this way, the spectral shapes as well as the duration of the selected records are kept constant during the process of record scaling. Therefore, the variability of responses at a given level of IM in IDA framework is mainly due to two matters: the first one is the concept known as the record-to-record variability and the next one is related to the level of IM at which responses are obtained. Record-to-record variability can include the dispersion of the structural responses for different applied GMs. As a result, we may have different structural responses when our structure is subjected to different ground motion records. Besides, the range of this response variability may get changed (or widened) when the intensity level of the applied GMs is changed (or increased) for different considered levels of IM. As a result, the dispersion of the damage measure given an IM can be due to both above-mentioned factors: the changed status of the IM level and the record-to-record variability.

In the proposed method described in this paper, levels of IM are not increased with the GM record scaling, but the IM is intensified while both the spectral shape and motion duration of the employed earthquake records get changed for all selected or required IMs. In other words, two other sources of variability are also added to the ones existed in the standard IDA procedure. The first one is the spectral shape of the records associated with a specific IM level. And the next source of variability incorporated into the proposed framework is the motion duration of the records employed at a given level of IM. Consequently, at each given level of IM, we have a set of records possessing two distinctive characteristics: a median duration and a unique acceleration spectrum.

On the way to consider the variability of motion duration of the produced GMs at a given level of IM, it is revealed that PGA (or SA) are nonlinearly correlated and convoluted with the significant duration parameter of the ground motions anticipated at a specific site. It means that an earthquake with a naturally generated PGA corresponds to a significant duration. Hence, the linear scaling procedure used in frameworks such as IDA may disturb the natural characteristics of the selected motions since the duration of the scaled ground motion is not altered and kept as it was before. Moreover, the relationship between PGA (or SA) and significant duration has found to be an exponential form composed of a region with a steady growth rate, along the horizontal or PGA axis, that gets longer as shorter rupture distances are taken to be at work. Typically, this region has short earthquake duration for all events simulated for a specific rupture distance. Since the fact the mentioned region gets widened enough when a small value of rupture distance is selected for simulation procedure of the scenario events, the PGA and motion duration seems to be somehow uncorrelated for such near-source earthquakes. This is in general agreement with the physical concept of the earthquakes in which rather shorter ground motion duration would be expected for near-source ground shakings. Therefore, this may let us think of record scaling as a legitimate way of changing the level of seismic intensity in case near-source ground motions are decided to be used in a dynamic procedure.

## 6. Summary and conclusion

A duration-consistent incremental dynamic analysis in which motion duration is incorporated in the structural analysis is proposed in this study. A simulation-based approach, which is verified for four real ground motions that are randomly selected from the NGA-West2 database, is employed to determine the median duration and the median acceleration spectra of earthquake motions at each intensity level. With having the median duration and the median acceleration spectra, artificial motions are generated at each intensity level. The functional relationship between duration and intensity level—which is required to estimate the median duration at each IM level—is investigated for the sites with different soil conditions and different rupture distances. It is demonstrated that exponential functions can get fitted to the simulated data and while the function is typically ascending, it can be expected that the median duration increases with the increase of intensity level. The sensitivity of the functions to rupture distance and soil conditions are also explored. Moreover, it has been found that the resulted shapes of acceleration spectra get altered at different intensity levels. The proposed duration-consistent incremental seismic assessment is used in nonlinear seismic assessment of two single degree of freedom structures—with and without a degrading behavior capability.

A standard or conventional IDA study has been selected to be as a benchmark to check the performance of the proposed incremental-based seismic framework. While the ground motions of the proposed method have inherently different duration and spectral shapes at each level of IM, for the IDA study, all motions have nearly constant durations and response spectra. This is because the motions picked for standard IDA are selected from one of those IM levels based on which ground motions of the proposed method are generated. It is revealed that considering the alteration and updating of duration and the shape of acceleration spectra of the ground motions at each intensity level can have substantial and meaningful impacts on structural responses of the nonlinear systems, which cannot be easily ignored in incremental-based seismic assessments. This is mainly due to the fact that incremental response curves, both for degrading and non-degrading SDOFs, derived from the proposed method show more nonlinearity compared to the standard IDA curve and demonstrate more softening at the higher IM levels.

In order to separately explore the variability of structural responses in terms of duration or spectral shape of motions, two other cases with new sets of ground motions are defined in addition to the artificial records we generated for the proposed method. In one case, only duration consistency is considered while the consistency of acceleration spectra is only included in the other one. This means that the first group of motions has a constant spectral shape— the one used for IDA study—but varied motion durations that have been already estimated by the proposed method at each level of seismic intensity. In the second group of ground motions, durations remain equal to the value of motion duration in the IDA study while the functions of the spectral shape at each level of seismic intensity are extracted from the simulation procedure. Compared to the IDA study, outputs of the dynamic analyses from ground motions in the first and second groups demonstrate that variability inclusion of motion duration or spectral shapes of the motions can cause their related response curves to get softened at the higher seismic intensity levels. However, the variability analysis shows that the most part of variability associated with the proposed method can be attributed to the alteration of spectral shape rather than duration. The softening effects associated with the variability of motion duration and alteration of spectral shapes of the motions are added to each other and get amplified in the proposed method that can synchronize duration to the amplitude-based IM, resulting in a flattened response curve or dynamic instability at the higher levels of seismic intensity.

## Acknowledgments

The authors are thankful to Dr. Kioumars Afshari for providing the data of NGA-West2 database along with his constructive comments which are gratefully appreciated. We would also like to thank Mr. Mirfarhadi for his constructive comments on the earlier and revised form of this paper.